\documentclass[conference]{IEEEtran}

\usepackage{mathtools}

\usepackage{mdwmath}
\usepackage{mdwtab}

\usepackage{array}

\usepackage{environ}
\usepackage[noend]{algpseudocode}
\usepackage{booktabs}
\usepackage{cite}
\usepackage{color}

\usepackage[pdftex]{graphicx}

\usepackage{url}

\usepackage{stfloats}

\usepackage{epsfig}

\usepackage{tabularx}
\usepackage{rotating}
\usepackage{hhline}
\usepackage{times}
\usepackage{comment}
\usepackage{amssymb}
\usepackage{amsthm}
\usepackage{pifont}
\usepackage{multirow}
\usepackage{footnote}
\usepackage{hhline}
\usepackage[bookmarks=false,colorlinks=true,citecolor=blue,linkcolor=blue,urlcolor=blue]{hyperref}
\usepackage{xspace} 
\usepackage{setspace}
\usepackage{bm}
\usepackage[export]{adjustbox}

\usepackage{microtype} 
\usepackage{diagbox}
\usepackage{listings}

\usepackage[caption=false,font=footnotesize]{subfig}

\microtypecontext{spacing=nonfrench}

\usepackage{xcolor}
\usepackage{soul}

\newcommand{\ie}{\emph{i.e., }}
\newcommand{\eg}{\emph{e.g., }}

\theoremstyle{definition}

\definecolor{darkgray}{rgb}{.4,.4,.4}
\definecolor{lightblue}{HTML}{CCFFFF}
\definecolor{lightgray}{HTML}{CCCCCC}
\definecolor{lightred}{HTML}{FF6666}
\definecolor{lightorange}{HTML}{FFCC99}

\def\sys{\textsc{XDA}\xspace}
\def\msvc{MSVC\xspace}

\newcommand{\etal}{\textit{et al.~}}

\DeclareMathOperator*{\argmin}{arg\,min}

\begin{document}
\title{\sys: Accurate, Robust Disassembly with\\Transfer Learning}


\makeatletter
\newcommand{\linebreakand}{%
  \end{@IEEEauthorhalign}
  \hfill\mbox{}\par
  \mbox{}\hfill\begin{@IEEEauthorhalign}
}
\makeatother

\author{\IEEEauthorblockN{Kexin Pei$^*$}
\IEEEauthorblockA{Columbia University\\
kpei@cs.columbia.edu}
\and
\IEEEauthorblockN{Jonas Guan$^*$}
\IEEEauthorblockA{University of Toronto\\
jonas@cs.toronto.edu}
\and
\IEEEauthorblockN{David Williams-King}
\IEEEauthorblockA{Columbia University\\
dwk@cs.columbia.edu}
\linebreakand
\IEEEauthorblockN{Junfeng Yang}
\IEEEauthorblockA{Columbia University\\
junfeng@cs.columbia.edu}
\and
\IEEEauthorblockN{Suman Jana}
\IEEEauthorblockA{Columbia University\\
suman@cs.columbia.edu}}

\date{}

\IEEEoverridecommandlockouts
\makeatletter\def\@IEEEpubidpullup{6.5\baselineskip}\makeatother
\IEEEpubid{\parbox{\columnwidth}{
    Network and Distributed Systems Security (NDSS) Symposium 2021\\
    21-24 February 2021, San Diego, CA, USA\\
    ISBN 1-891562-66-5\\
    https://dx.doi.org/10.14722/ndss.2021.23112\\
    www.ndss-symposium.org
}
\hspace{\columnsep}\makebox[\columnwidth]{}}

\maketitle

\makeatletter
\def\blfootnote{\xdef\@thefnmark{}\@footnotetext}
\makeatother

\blfootnote{$^*$Equal contribution, order decided by coin flip}

\begin{abstract}

Accurate and robust disassembly of stripped binaries is challenging. The root of the difficulty is that high-level structures, such as instruction and function boundaries, are absent in stripped binaries and must be recovered based on incomplete information. Current disassembly approaches rely on heuristics or simple pattern matching to approximate the recovery, but these methods are often inaccurate and brittle, especially across different compiler optimizations. 

We present \sys, a transfer-learning-based disassembly framework that learns different contextual dependencies present in machine code and transfers this knowledge for accurate and robust disassembly. We design a self-supervised learning task motivated by masked Language Modeling to learn interactions among byte sequences in binaries. The outputs from this task are byte embeddings that encode sophisticated contextual dependencies between input binaries' byte tokens, which can then be finetuned for downstream disassembly tasks.

We evaluate \sys's performance on two disassembly tasks, recovering function boundaries and assembly instructions, on a collection of 3,121 binaries taken from SPEC CPU2017, SPEC CPU2006, and the BAP corpus. The binaries are compiled by GCC, ICC, and \msvc on x86/x64 Windows and Linux platforms over 4 optimization levels. \sys achieves 99.0\% and 99.7\% F1 score at recovering function boundaries and instructions, respectively, surpassing the previous state-of-the-art on both tasks. It also maintains speed on par with the fastest ML-based approach and is up to 38$\times$ faster than hand-written disassemblers like IDA Pro. We release the code of \sys at \url{https://github.com/CUMLSec/XDA}.

\end{abstract}

\section{Introduction}
\label{sec:intro}

Disassembly is the backbone of many binary analysis tasks, such as malware analysis, reverse engineering, retrofitting control-flow integrity defenses, binary rewriting, and binary instrumentation~\cite{dolan2015repeatable, snow2013just, wang2017ramblr, shoshitaishvili2016state, peng2018t}. Binary analysis relies on disassembly to recover higher-level constructs such as assembly instructions and function boundaries from machine code.

Disassembly is difficult because high-level information from symbol tables and source code are absent in stripped binaries. This information is either discarded during compilation or stripped before distribution to decrease program size or deter reverse engineering. Therefore, disassemblers must approximate the recovery of higher-level constructs from incomplete information. Complex assembly constructs such as inline data and tail calls, further complicate disassembly.

Traditional approaches to this problem rely on handcrafted heuristics to guide the recovery, but these methods are inaccurate and brittle~\cite{wartell2011differentiating, bao2014byteweight, andriesse2017compiler}. For example, many popular disassemblers, like IDA~Pro and Ghidra, recover assembly instructions by recursively following \emph{direct} control flow transfers (\eg \texttt{call 0x2ed4}). Similarly, they recover function boundaries by looking for \emph{known instruction patterns}. These heuristics-based methods are insufficient -- prior research has shown that IDA~Pro misidentifies up to 58\% of function boundaries~\cite{bao2014byteweight} and 4\% of assembly instructions~\cite{andriesse2016depth} in optimized binaries. Furthermore, many heuristics for detecting functions require continual manual maintenance of large databases to adapt to code and compiler changes~\cite{bao2014byteweight}:
IDA~Pro's current database for identifying common library functions is over 41MB in size, and Ghidra's is over 179MB.

Previous research has explored using Machine Learning (ML) to address these challenges. The resulting models surpassed the accuracy of traditional disassemblers at both recovering assembly instructions~\cite{wartell2011differentiating} and function boundaries~\cite{bao2014byteweight, shin2015recognizing}, and are easier to maintain.
However, current ML-based approaches still face two critical challenges: 

\noindent\textbf{Accuracy.}
Recent research has shown that previous ML-based methods are likely not as accurate as reported~\cite{andriesse2017compiler}. 
This is because their accuracies were inadvertently measured on testing data with significant overlap with training data, and therefore their performance may not generalize to real-world binaries.
More concretely, Andriesse~\etal\cite{andriesse2017compiler} showed that the F1 score of ByteWeight~\cite{bao2014byteweight} degrades from the reported 97\% to 65\% when evaluated on a dataset without overlap.

\noindent\textbf{Robustness.} The current most accurate ML approach for recovering function boundaries, bidirectional Recurrent Neural Net (bi-RNN)~\cite{shin2015recognizing}, is not robust to compiler optimization changes (see Section~\ref{subsec:robust_analysis}).
Similar concerns have also been raised by Wang~\etal\cite{wang2017semantics}, who proposed a more robust method by combining ML with symbolic execution. However, this method does not scale well to larger binaries.

\begin{figure*}[!t]
\centering
\includegraphics[width=0.98\linewidth]{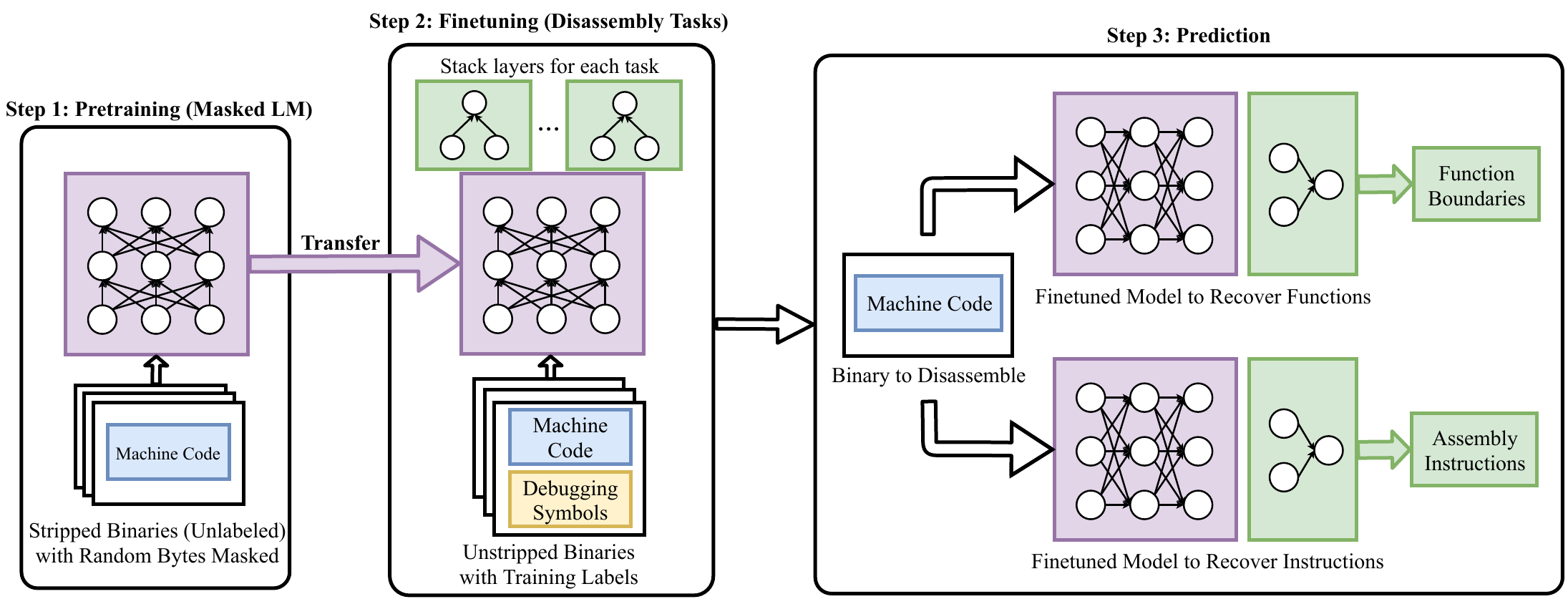}

\caption{
The workflow of \sys. We first pretrain the model with the masked LM task, and then finetune the pretrained model with new stacked NN layers for disassembly tasks. Finetuning updates both the pretrained model and the stacked layers. During inference, the finetuned models takes the binary and output disassembly information, depending on the finetuned tasks.}

\label{fig:general}
\end{figure*}

In this paper, we present \sys (Xfer-learning DisAssembler), a new ML-based disassembly framework that uses \emph{transfer learning} to address these challenges. We train and evaluate \sys on 3,121 binaries, and separate the testing and training data such that they have minimal overlap (less than 3\%). We show that \sys outperforms all state-of-the-art tools in accuracy at recovering function boundaries and assembly instructions, and is robust to changes in compiler optimization. Additionally, \sys's speed is on par with the fastest tools.

Our key insight is to teach the ML model general dependencies between bytes in machine code before training it to perform specific disassembly tasks -- we \emph{transfer} its learned knowledge of these byte dependencies to tackling disassembly. 
Intuitively, we first teach our model to read and gain a basic understanding of machine code and then teach it to solve a disassembly task.
We achieve this by splitting our training into two stages, as shown in Figure~\ref{fig:general}. In the first stage, we pretrain our model using masked Language Modeling (masked LM) to teach it the byte dependencies of x86/x64 machine code. In the second stage, we finetune our model to leverage its knowledge of byte dependencies to solve a specific disassembly task accurately and robustly.

The masked LM task asks the model to predict randomly masked bytes, which requires the model to produce missing bytes given the context. This design compels the model to learn dependencies between the masked byte and surrounding bytes, teaching the model a basic understanding of machine code semantics. 
As we will show in Section~\ref{sec:eval}, this understanding improves not only the model's accuracy but also its robustness to compiler and optimization changes. In contrast, previous ML approaches rely on superficial patterns, such as function prologues, which limits their accuracy and robustness when such patterns are absent or changed. An additional benefit is that the masked byte prediction task does not require labeled data. Therefore, \sys can be further trained and improved using stripped binaries found in the wild.

We give some intuition on why the masked LM task is helpful for disassembly. Consider a function that uses \texttt{0x28} bytes of stack space for local variables. To allocate this space, the function uses \texttt{sub rsp,0x28} to decrement the stack register by \texttt{0x28} bytes. Now assume the byte \texttt{0x28} is masked, like so: \texttt{sub rsp,??}. To predict the value of this masked byte, our pretrained model learns to search for the corresponding stack deallocation instruction (\texttt{add rsp,0x28}) and read its value (see Section~\ref{subsec:visualization} for more details). Since stack allocation/deallocation often occurs near the start/end of a function, the knowledge of this dependency is a helpful feature for identifying function boundaries. Such scenarios that naturally occur in predicting randomly masked bytes, in sum, teaches the model a powerful, general understanding of machine code semantics that helps disassembly.

To evaluate \sys's performance, we test it on 3,121 Linux and Windows x86/x64 binaries taken from the SPEC CPU2017~\cite{spec2006cpu}, SPEC CPU2006~\cite{spec2017cpu} benchmark suites and the BAP corpus~\cite{bao2014byteweight}. These binaries are compiled using the GNU Compiler Collection (GCC), Intel C++ Compiler (ICC), and Microsoft Visual C++ (\msvc) over 4 optimization levels (\texttt{O0-O3} for GCC and ICC,  \texttt{O1, O2, Od, Ox} for \msvc). We choose two popular disassembly tasks, recovering function boundaries and assembly instructions, to test our model's performance, but \sys can be easily finetuned for other disassembly and binary analysis tasks.
Across these binaries, \sys achieves 99\% F1 score at recovering function boundaries, 17.2\% higher than the second-best tool. \sys also achieves a 99.7\% F1 score at recovering assembly instructions. Furthermore, \sys's underlying neural architecture is highly parallelizable and efficient, running up to 38$\times$ faster than hand-written disassemblers like IDA Pro.

We also evaluate the robustness of \sys to the changes of compiler optimizations. \sys achieves at least 98.5\% F1 score at recovering function boundaries in highly optimized binaries, even when it is only finetuned on unoptimized binaries.

We make the following contributions.

\begin{itemize}
\item 
We propose a new approach to disassembly using a two-step transfer learning paradigm: we first pretrain the model to teach it a basic understanding of machine code, then finetune it to solve disassembly tasks.

\item 
We demonstrate that masked LM is an effective pretraining task for disassembly because it compels the model to learn machine code semantics (Section~\ref{sec:overview}). We then show how this semantic knowledge can be leveraged in finetuning to accurately and robustly solve two popular disassembly tasks, recovering function boundaries and assembly instructions.

\item
We implement \sys and evaluate it on a collection of x86/x64 binaries from the SPEC CPU2017, SPEC CPU2006, and BAP dataset on both the Windows and Linux platforms (Sections~\ref{sec:impl} and  \ref{sec:eval}). The binaries are compiled by GCC, ICC, and MSVC over 4 optimization levels.
Even when the pretraining, finetuning, and testing datasets are strictly separated, \sys achieves an average F1 score of 99.0\% at recovering function boundaries, outperforming the state-of-the-art by 17.2\%, and an average F1 score of 99.7\% at recovering assembly instructions.
We open-source our implementation at \url{https://github.com/CUMLSec/XDA}.

\end{itemize}

\section{Background}
\label{sec:background}

We briefly overview the two disassembly tasks we tackle, recovering function boundaries and assembly instructions, two fundamental building blocks in binary analysis research~\cite{andriesse2016depth}.
We then describe their challenges faced by existing tools using real-world examples.

The definition of disassembly can be ambiguous, as some authors refer only to the process of recovering instructions from binaries~\cite{miller2019probabilistic, bauman2018superset, wartell2011differentiating}. 
We adopt the more inclusive interpretation defined by Andriesse~\etal\cite{andriesse2016depth, andriesse2017compiler}:
Disassembly is the process of recovering higher-level program constructs lost during compilation, such as assembly instructions, function boundaries, function signatures, and control flow graphs.

Most disassembly tools recover these constructs via static analysis, because dynamic analysis has large runtime overheads and struggles to achieve high completeness in complex programs. Therefore, we also limit our methods to operate on information acquirable without executing the program.

\subsection{Recovering Function Boundaries}
The task of recovering function boundaries consists of identifying the matched start and end addresses of each function in a stripped binary.
We inherit the formal definition of this task by Bao \etal\cite{bao2014byteweight}: given a binary that contains a set of functions $F = \{f_{1}, f_{2}, ..., f_{n}\}$, recover a set of function start and end address pairs $S = \{(s_{1}, e_{1}), (s_{2}, e_{2}), ..., (s_{n}, e_{n})\}$ such that for all $f_{i} \in F,\ s_{i}$ is the address of the first byte of $f_{i}$ and $e_{i}$ marks the end of $f_{i}$ (the first byte not contained in $f_{i}$).

Recovering function boundaries from stripped binaries is one of the most challenging disassembly tasks~\cite{andriesse2016depth, bao2014byteweight, meng2016binary, harris2005practical, vigna2007static, shoshitaishvili2015firmalice, shoshitaishvili2016state, di2017rev}. Functions are source-level constructs but decay to simple control-flow transfer at the machine code level, and symbol tables containing function information are removed. Compiler optimizations exacerbate the problem by removing indicative structures such as function prologues. For disassemblers operating on recovered assembly instructions instead of machine code, an additional concern is functions whose instructions are not recovered cannot be identified.

\subsection{Recovering Assembly Instructions}
The task of recovering assembly instructions consists of identifying the bounds of each individual assembly instruction within the code sections of a stripped binary.

The main difficulty of recovering assembly instructions comes from distinguishing inline data from code via static analysis~\cite{miller2019probabilistic}. Architectures like x86 and x64 have variable-length instructions and do not make syntactic distinctions between inline data and code~\cite{bauman2018superset, wartell2011differentiating}. 
However, compilers such as \msvc interleave data with code for code efficiency. 
In the worst case, the only difference between data and code at the machine code level is that code bytes are reachable at runtime. Therefore, perfectly recovering assembly instructions from x86 and x64 binaries is undecidable~\cite{wartell2011differentiating, wartell2014shingled}. Mistakenly parsing a data byte as the start of a multi-byte instruction can desynchronize the disassembled instruction stream alignment, propagating the error forward~\cite{andriesse2016depth}.

\begin{figure*}[!t]
\centering

\includegraphics[width=0.9\linewidth]{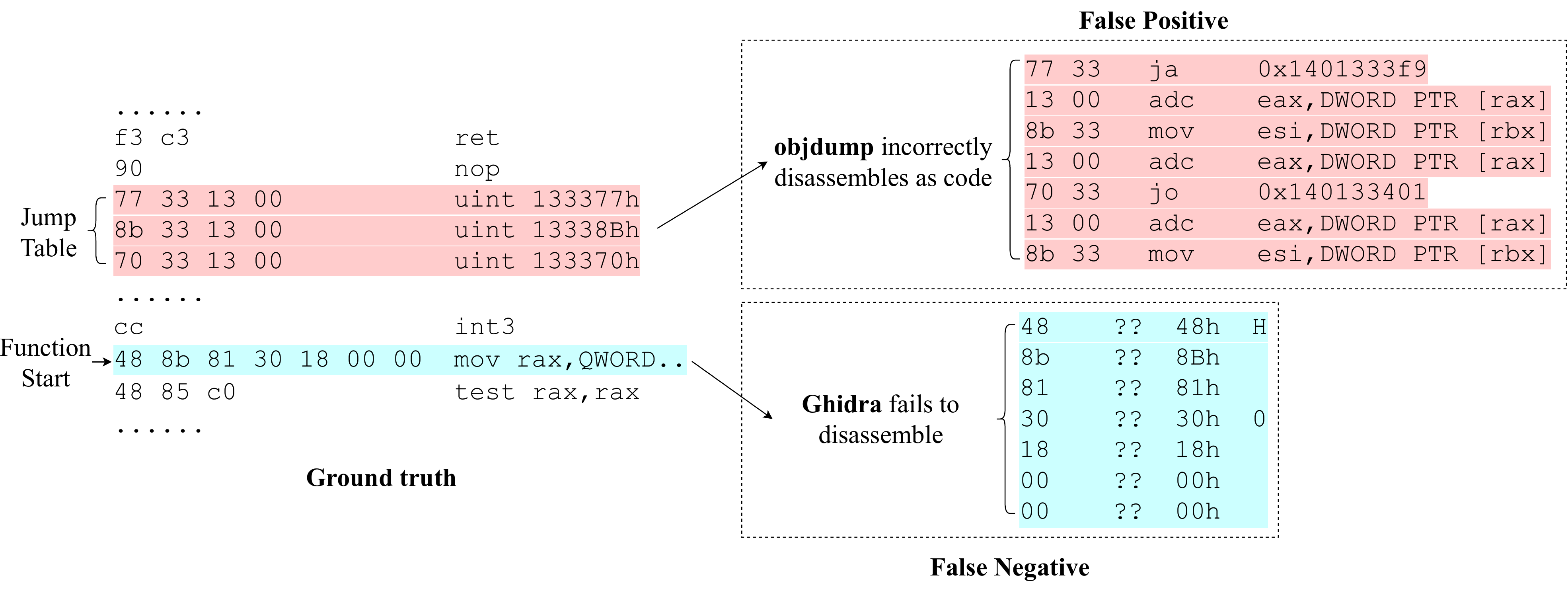}

\caption{Real-world example that fails objdump and Ghidra. 
The ground truth assembly instructions include (1) a \emph{jump table in the code section} that objdump cannot identify, and (2) a function reached via an \emph{indirect} control flow transfer that Ghidra cannot statically determine.}
\label{fig:objdump_ghidra_fail}
\end{figure*}

\begin{figure}[!t]
\centering

\includegraphics[width=0.85\linewidth]{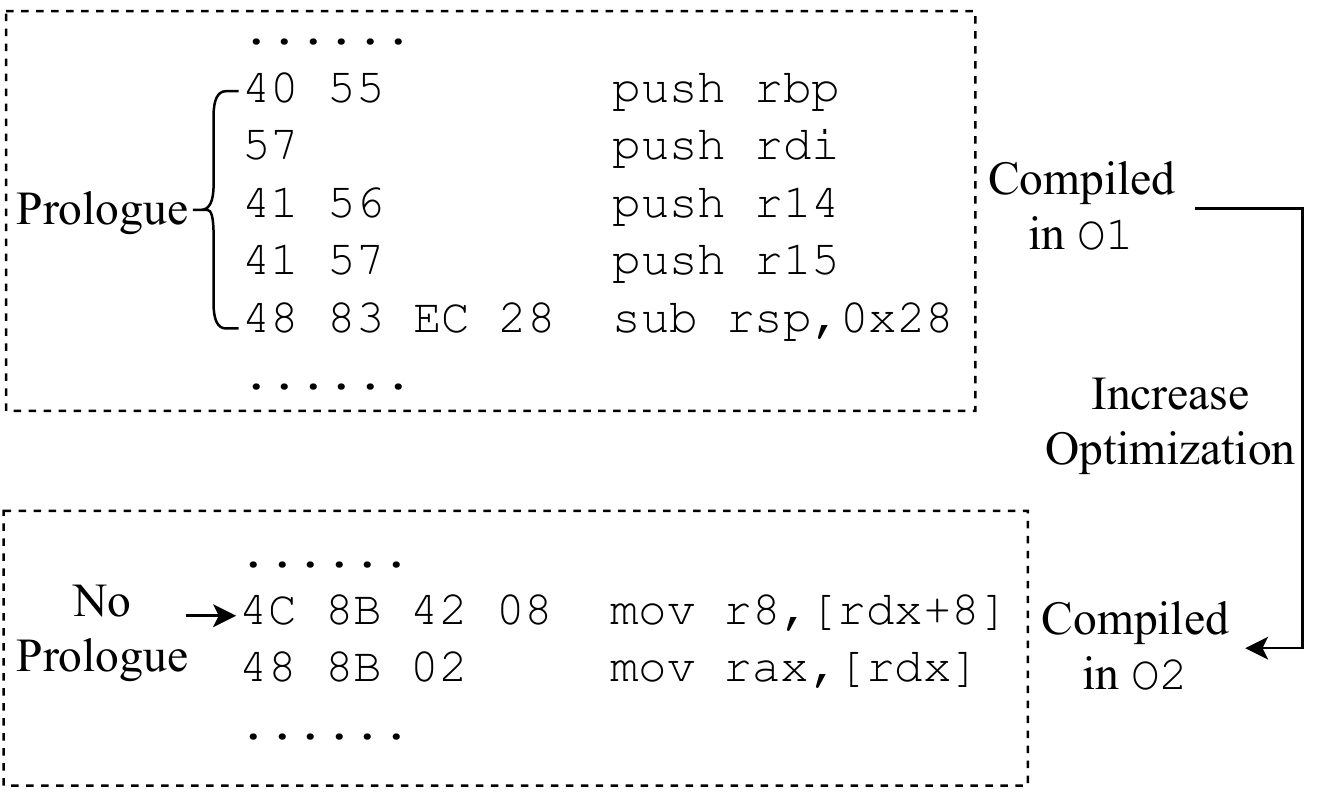}

\caption{Both IDA~Pro and bi-RNN recognized this function when it was not optimized (\texttt{O1}), but failed to recognize it in higher optimization (\texttt{O2}) when the function prologue is optimized away.}
\label{fig:ida_birnn_fail}
\end{figure}

\subsection{Challenging Cases}

We use code snippets from Vim 8.2 compiled by \msvc 2019 to demonstrate some challenges faced by heuristic-based approaches. This serves primarily as a motivating example. For a more thorough listing of challenging disassembly cases, we refer interested readers to Andriesse~\etal\cite{andriesse2016depth}.

Traditional disassemblers rely on heuristics to recover function boundaries and assembly instructions, requiring data such as control flow information and function prologues/epilogues, which restricts their accuracy and robustness.
For example, Figure~\ref{fig:objdump_ghidra_fail} shows the linear disassembler objdump and the recursive traversal disassembler Ghidra failing to disassemble a section of code. objdump misinterpreted a jump table as instructions, whereas Ghidra missed an entire function of code. Similarly, in Figure~\ref{fig:ida_birnn_fail} IDA~Pro and bi-RNN fail to recognize a function when the compiler optimization is increased, removing the function prologue.



In contrast, \sys correctly recovers all instructions and functions in this code. The key to this success is \sys's transfer learning approach, which enables it to capture general patterns that match real program dependencies and semantics. We provide the intuition behind this in the following section.


\section{Overview of Our Approach}
\label{sec:overview}

We have shown \sys's workflow of pretraining and finetuning in Figure~\ref{fig:general}.
In this section, we describe how the pretraining task, masked LM, compels the model to learn machine code dependencies that could support many downstream disassembly tasks.
Although formally proving such claims remains an open question~\cite{arora2019theoretical}, we sketch two intuitive cases below to motivate why predicting masked bytes teaches the model machine code semantics that helps disassembly. We include the more complete details in Sections~\ref{subsec:probe_case}.


Recall the definition of the masked LM pretraining task: given a byte sequence, we mask out some random bytes, and train the model to predict the masked bytes using the non-masked bytes. The non-masked bytes are known as the contextual information. 
We use two examples to show that pretraining with masked LM allows a model to thoroughly understand properties of machine code, aiding recovery of function boundaries and assembly instructions.


\noindent\textbf{Masking local variable allocation.}
Consider the code snippet in Figure~\ref{fig:ida_birnn_fail}, where we mask four bytes (\texttt{48 83 ec 28}), representing \texttt{sub rsp,0x28} that decrements the stack pointer to allocate 40 bytes for local variables. 
Unlike the example described in Section~\ref{sec:intro}, the model does not have access to the corresponding stack deallocation instruction.
We found that our model still correctly predicts the four bytes with high confidence, which implies the model understands the \emph{semantics of stack allocation and can deduce the space required by local variables} by looking at the surrounding bytes in the function body. 
Besides, the understanding of such semantics can \emph{significantly help in recovering function boundaries}, because the function prologue strongly indicates a function start.
Section~\ref{subsec:pretrain} shows pretraining with such tasks improves the results by 50\% at recovering function boundaries.

\noindent\textbf{Masking jump table entries.} Now consider the jump table example in Figure~\ref{fig:objdump_ghidra_fail}, where we mask out the jump table entry \texttt{77 33 13 00}.
We find that our model correctly predicts the four bytes with high confidence, which indicates it learns to \emph{distinguish inline data from code}. 
For example, if it treats the masked bytes as instructions (like those disassembled by objdump shown on the right), it is highly unlikely for the model to predict \texttt{77 33} (the \texttt{ja} instruction) right after the instructions \texttt{ret; nop}. 
The reason is that compiler-generated byte code (\sys's training data) rarely places a conditional jump like \texttt{ja}, which depends on flags set by \texttt{cmp} or \texttt{test}, after an \texttt{nop}. 
Therefore, the model should understand the masked bytes are part of the inline data section, which should be similar to the following (as opposed to the preceding) bytes corresponding to other jump table entries.

To confirm that the model is not simply memorizing the byte order in \texttt{Vim}, we remove all \texttt{Vim} binaries (shown in Figures~\ref{fig:objdump_ghidra_fail} and~\ref{fig:ida_birnn_fail}) from the training set. 
Therefore, the model's high accuracy strongly indicates that the semantics described above is captured in the pretraining process.

\section{Threat Model}
\label{sec:threat_model}

\noindent\textbf{Robustness.} We focus on binaries generated by standard compilers. Like other disassemblers, we do not aim to be robust in the presence of arbitrarily obfuscated code. Instead, we aim to be robust against compiler changes, which often occur due to improvements in optimization. In Section~\ref{subsec:robust_analysis}, we evaluate the robustness of our model by testing it on binaries compiled on higher optimization levels than the model is trained on.

\noindent\textbf{False positives and negatives.} Like most disassemblers, we assume a small number of false positives and negatives can be tolerated. Certain use cases in binary rewriting require recovered instructions to contain strictly no false positives or no false negatives; we do not enforce these requirements.

\section{Methodology}
\label{sec:method}

We describe the design specifics of \sys, including the key components of the architecture and workflow of pretraining and finetuning. We then elaborate on the technical details in the following subsections.

\noindent\textbf{General design.}
Figure~\ref{fig:arch} shows the simplified architecture of \sys and bi-RNN~\cite{shin2015recognizing} and their example input-output. 
It highlights \sys's two key design differences with a typical ML-based technique, bi-RNN~\cite{shin2015recognizing}.
The first is the employment of a pretraining task. 
As discussed in Section~\ref{sec:overview}, this design encourages the model to learn machine code contextual dependencies helpful for many binary analysis tasks.  

Second, we employ \emph{self-attention} layers~\cite{vaswani2017attention} in the model to compute information flow between every pair of bytes. 
Specifically, we follow the encoder architecture in Transformer~\cite{vaswani2017attention} with several modifications (Section~\ref{subsec:pretrain_method}).
This design is shown more capable than the sequential connections in RNNs in capturing long-range dependencies between distant bytes. As we will show in Section~\ref{sec:case}, the model often needs to learn and understand long-range dependencies in the binaries to recover functions boundaries and instructions accurately.

\begin{figure}[!t]
\centering

\subfloat[\sys architecture]{
\includegraphics[width=0.46\linewidth]{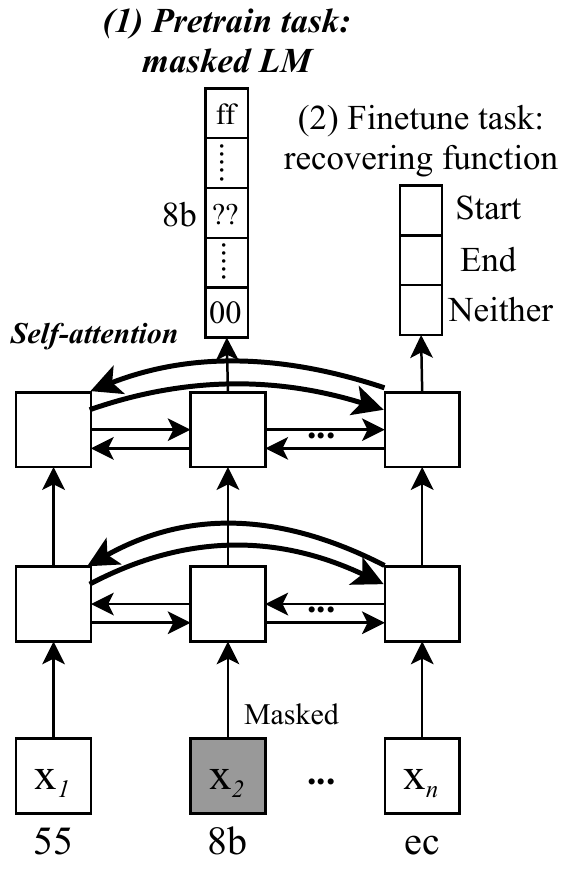}
\label{subfig:bert}}
\subfloat[bi-RNN architecture~\cite{shin2015recognizing}]{
\includegraphics[width=0.46\linewidth]{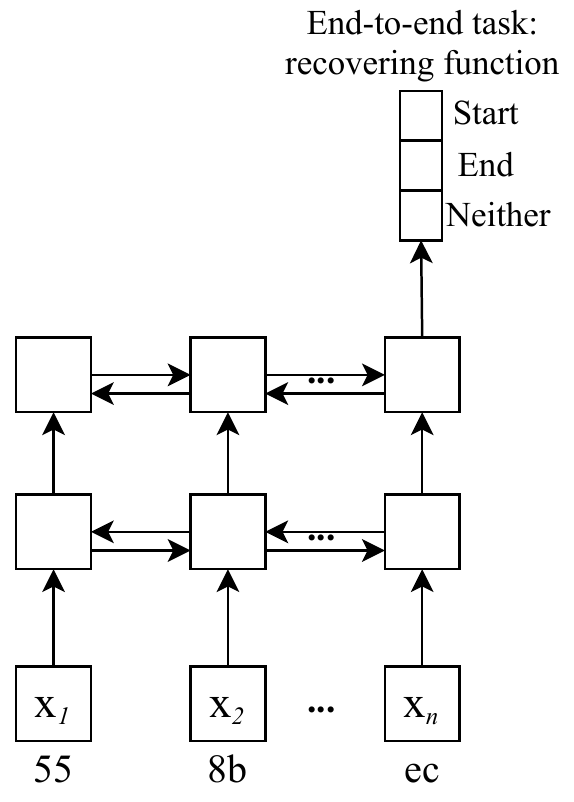}
\label{subfig:birnn}}

\caption{
Architectural differences between \sys and bi-RNN used by Shin \etal\cite{shin2015recognizing}. Two key distinguishing components of \sys are: (1) the pretraining (masked LM) steps, and (2) self-attention connections between arbitrary tokens instead of only between neighbors.}
\label{fig:arch}
\end{figure}

\noindent\textbf{Input representation.} Formally, we define the input $x$ as a sequence of byte tokens of size $n$: $x=$\{\texttt{0x00}, ..., \texttt{0xff}\}$^n$. 
Each input byte $x_i\in x$ is represented as a one-hot encoded vector (\eg \texttt{a3} is encoded as a 256-dimensional vector with all 0s but single 1 at position 163). 
Besides 256 possible byte values, the input vocabulary has 5 additional reserved tokens (\ie padding \texttt{<PAD>}, start-of-sequence \texttt{<S>} and end-of-sequence \texttt{</S>}, unknown \texttt{<UNK>}, and mask \texttt{<MASK>}).
Also, note that here we do not put constraints on $n$. For example, the byte sequences can span multiple binary programs, or include only a subset of byte sequence within a single binary. 

\noindent\textbf{Pretraining task.}
We pretrain the model using the masked LM objective.
Formally, for each byte sequence $x_1, ...,x_n$ in a given training set, we mask out some percentage of byte tokens in each sequence randomly (see Section~\ref{subsec:mask} for details). 
The masked byte tokens are replaced by the mask tokens (\texttt{<MASK>}).

Let $mask(x_i)$ denote the output of applying the mask on the $i$-th byte in $x$ and $mpos$ a set of masked byte's positions in $x$. 
The model to be pretrained, $f_p$, will take as input the byte sequence with random bytes masked: $(x_1,..., mask(x_i),...,x_n), i\in mpos$, and predicts the byte values of the masked tokens: $\{\hat{x}_i|i\in mpos\}=f_p(x_1,..., mask(x_i),...,x_n)$.
Let $f_p$ be parameterized by $\theta$ ($f_p(-;\theta)$), the objective of training $f_p$ is thus to search for $\theta$ that minimizes the cross entropy loss between the predicted masked bytes and the actual bytes. For ease of exposition, we omit summation over all samples in the training set.

\begin{equation}
\label{eq:pretrain}
\centering
    \argmin_{\theta} \sum\limits_{i=1}^{|mpos|} -x_i\log(\hat{x}_i)
\end{equation}

While the pretrained model's input and output are one-hot encoded byte tokens, all of its intermediate layers operate on the embedding vectors. 
For each byte $x_i$, let $E_{1,i}$ denote its embedding after $1$-st layer, the embeddings produced by the $l$-th layer of the model is $E_{l}=(E_{l,1},...,E_{l,n})$. 
Let layer $l$ be the $f_p$'s last layer; $E_{l}$ is not the actual outputs of $f_p$, as $f_p$ will take $E_{l}$ and stack a classification layer (\eg softmax) to predict the value of the byte tokens.
Given the produced embeddings $E_l$, we define finetuning tasks in the following.

\noindent\textbf{Finetuning tasks.}
Given a sequence of embedding vectors produced by the pretrained model $E_{l}=(E_{l,1},E_{l,2},...,E_{l,n})$, the corresponding ground truth $y$ in our finetuning task is a sequence of labels of same length $n$: $\{y_i | y_i\in C_y\}^n$, where $C_y$ denote the set of all possible class labels of $y$. 
Each byte embedding $E_{l,i}$ will be mapped to a label $y_i$.

Let $f_t$ denote the model to be finetuned. It takes as input each byte embedding $E_{l,i}$, and predicts the label $\hat{y_i}$, where $\hat{y_i}=f_t(E_{l,i})$.
Now let $f_t$ be parameterized by $\theta$: $f_t(E_{l,i};\theta)$, the objective of training $f_t$ is thus to search for $\theta$ that minimizes the cross entropy between the predicted labels of all byte embeddings and their actual labels:

\begin{equation}
\label{eq:finetune}
\centering
    \argmin_{\theta} \sum\limits_{i=1}^n -y_i\log(\hat{y_i})
\end{equation}

As both $f_p$ and $f_t$ in our setting are neural networks, solving Equations~\ref{eq:pretrain} and~\ref{eq:finetune} can be efficiently guided by gradient descent via backpropagation.
The gradient can flow through both $\theta$s in $f_p$ and $f_t$ during finetuning, as $f_p$ is also part of the computation of $\hat{y}^{(i)}_j$. Thus, the pretrained parameters in $f_p$ also get updated during finetuning to adjust for specific downstream tasks.


As a concrete example, consider recovering function boundaries. The possible class labels $C_y=\{S, E, N\}$ denote the function start ($S$), end ($E$), and neither ($N$). $f_t$ takes each byte embedding as input and produces the probability distributions of three possible labels, where the predicted label will be the one with the highest probability. This is illustrated in Figure~\ref{subfig:bert}.


\subsection{Masked Language Model on Binaries}
\label{subsec:pretrain_method}

As illustrated in Figure~\ref{subfig:bert}, the forward pass within the \sys's architecture consists of the following steps. 

First, \sys embeds the one-hot vectors of each input byte (both masked and not masked) as a fixed-dimension embedding vector.
What is not shown in Figure~\ref{subfig:bert} is that \sys also computes the \emph{positional embeddings}, which encode the position of each byte within the sequence as another vector that has the same size with the byte embeddings.
The rationale of using the positional embeddings is to assign distinctive meanings to the same byte tokens appearing in different locations within the sequence. Note that in recurrent-based networks, such spacial relationship is naturally encoded as the information flow follows the order in the input sequence~\cite{vaswani2017attention}.
Two embeddings (\ie byte embeddings and positional embeddings) are then combined as the actual embeddings of input for the next layer.

\sys then employs \emph{multi-head self-attention} to update the embeddings. At each self-attention layer, each embedding combines itself with a weighted sum of all other embeddings, where the attention strength determines the weight.

Finally, \sys predicts the byte values of masked bytes based on its updated embeddings in the last layer. To this end, \sys stacks a 2-layer fully-connected network with the output dimension equal to the input vocabulary size. The updated embedding here is known as \emph{contextualized embedding}~\cite{mikolov2013efficient, mikolov2013distributed, pennington2014glove, peters2018elmo, devlin2018bert} as it takes into account the information of other byte tokens within the sequence.

In the following, we elaborate on the key concepts described above. We will use the same notations defined in Section~\ref{sec:method} wherever they are applicable.

\noindent\textbf{Byte position embeddings.}
The position of each input byte is critical for inferring binary semantics. Unlike natural language, where swapping two words can roughly preserve the same semantic meaning, swapping two bytes can significantly change the instructions.
Therefore, we use the \emph{learned positional embedding} $E_{pos}$ (embedding matrix itself is learnable) and stack a feedforward network $F$ with one hidden layer to combine $E_{pos}$ with byte embeddings $E_{byte}$.
Specifically, let $E_{byte}(x)$ denote applying the embedding to the one-hot encoded byte token $x_i$, and let $E_{pos}(i)$ denote applying the learned positional embedding to $x_i$'s position $i$, we have: $ E_{1,i} = F(concat(E_{byte}(x_i),E_{pos}(i)))$.
As defined in Section~\ref{sec:method}, $E_{1,i}$ here refers to the embedding of $x_i$ in the 1st attention layer.

\noindent\textbf{Multi-head self-attention.}
Given the embeddings of all bytes in $k$-th layer, $(E_{k,1},...,E_{k,n})$, the core operations of a single attention head in self-attention layer is to update each embedding by the following steps.

First, each embedding $E_{k,i}$ will get mapped to three values following query-key-value computation: $query_i=f_{query}(W_{query};E_{k,i})$, $key_i=f_{key}(W_{key};E_{k,i})$, $value_i=f_{value}(W_{value};E_{k,i})$. Here $f_{query}$, $f_{key}$, and $f_{value}$ are affine transformation functions (\ie a fully-connected layer) parameterized by $W_{query}$, $W_{key}$, and $W_{value}$, respectively.

Each embedding $E_{k,i}$ then computes attention score $s_{ij}$ with all other embeddings $E_{k,\{j|j\neq i\}}$ by taking the dot product between the $E_{k,i}$'s query $query_i$ and $E_{k,j}$'s key $key_j$: $s_{ij}=query_i\cdot key_j$. 
Intuitively, attention scores $s$ for all pairs will end up as a square matrix, where each cell $s_{ij}$ indicates how much attention $E_{k,i}$ should pay to $E_{k,j}$ when updating itself. 
We then divide every row of $s$ by $\sqrt{d_{emb}}$ (the dimension of the embedding vectors) and scale it by softmax to ensure them sum up to 1: $s'_{ij} = \frac{\exp(s_{ij})}{\sum^n_{j=1}\exp(s_{ij})}$
The scaled attention score $s'_{ij}$ will be multiplied with $value_j$ and summed up: $E^h_{k+1,i} = \sum^n_{j=1} s'_{ij} v_j$

Here $h$ in $E^h_{k+1,i}$ denote the updated embeddings belong to attention head $h$. Assume we have total $H$ attention heads, the updated embeddings will finally go through an 2-layer feedforward network $f_{out}$ parameterized by $W_{out}$ with skip connections~\cite{he2016deep} to update embeddings from all heads: $E_{k+1,i} = f_{out}(concat(E^0_{k+1,i},...,E^H_{k+1,i});W_{out})$.

\noindent\textbf{Contextualized embeddings.}
The multi-layer attention mechanism updates its embeddings iteratively up to the last layer. The embeddings at each self-attention layer are known as contextualized (or context-aware) embeddings. 
One significant feature of contextualized embeddings is that the underlying meaning of each embedding depends on the other tokens in the input sequence. So the embeddings for the same byte token can be different if the byte is in a different context (surrounded by different bytes).
This is in contrast with static embeddings (\eg word2vec~\cite{mikolov2013distributed}) commonly used by other related works (\eg learn binary embeddings~\cite{duandeepbindiff, ding2019asm2vec}), where a byte token is always assigned to the same fixed embedding regardless of the changed context (\ie surrounding bytes).


\subsection{Distilling the Learned Semantics}
\label{subsec:finetune_method}

The examples in Section~\ref{sec:overview} motivate how pretraining on masked LM task encourages learning semantics.
However, they are all implicitly embedded in the weight parameters of the underlying neural network. 
To distill the learned semantics for different downstream tasks, we leverage the contextualized embeddings produced by the pretrained model's last layer and stack a simple 2-layer multi-layer perceptron (MLP) for prediction. 
Explicitly, for each embedding $E_{l,i}$ in the last layer, we fix the network to perform the following computation: $MLP(E_{l,i}) = softmax(tanh(E_{l,i}\cdot W_1)\cdot W_2)$.
Here $W_1\in \mathbb{R}^{d_{emb}\times d_{emb}}$ and $W_2\in \mathbb{R}^{d_{emb}\times |C_y|}$ where $|C_y|$ is the number of output classes (defined in Section~\ref{sec:method}). 
Note that we apply $MLP$ to each byte embedding. Therefore, the output of $MLP$ is of shape $\mathbb{R}^{n\times |C_y|}$, where $n$ is the length of the input sequence.


\subsection{Masking Input Bytes}
\label{subsec:mask}

For each input sequence in pretraining, we choose 20\% random bytes to mask.
Among the chosen bytes, we select 50\% of them to be replaced by the special token \texttt{<MASK>}. For the remaining 50\% of the masked bytes, we replace with random bytes in the vocabulary \{\texttt{0x00}, ..., \texttt{0xff}\}. 
The reason of not replacing all chosen bytes by \texttt{<MASK>} is that, in the finetuning task, there is no such token. Therefore, we try to prevent the model from discovering any spurious meanings of the \texttt{<MASK>} token itself, but encourage the model to focus on using the context to predict the masked tokens.

For the same input sequences at different epoch, we \emph{re-randomize} the bytes to mask instead of fixing the same set of masked bytes throughout different epochs. Therefore, we implement a dynamic masking scheme and do not apply masking in the data preprocessing stage.

\section{Implementation and Experimental Setup}
\label{sec:impl}

We build the learning module of \sys in PyTorch 1.4.0 with CUDA 10.1 and CUDNN 7.6.3. 
We implement the self-attention architecture using Fairseq toolkit~\cite{ott2019fairseq}.
We run the training and inference of our models on a Linux server running Ubuntu 18.04, with an Intel Xeon E5-2623 at 2.60GHz with 16 virtual cores including hyperthreading, 256GB RAM, and 3 Nvidia GTX 1080-Ti GPUs. 

\noindent\textbf{Datasets.}
Table~\ref{tab:dataset} summarizes the datasets we use for training and evaluating \sys. 
The first dataset is SPEC CPU2017~\cite{spec2017cpu}, the updated version of SPEC CPU benchmarks. It includes 39 C/C++/Fortran compute-intensive programs. 
We compile each program using 2 compilers, GCC-9.2 and \msvc 2019 on Linux and Windows, respectively. Each compiler compiles the program on 2 ISAs (x86 and x64) with respective 4 optimization levels (\eg \texttt{O1}, \texttt{O2}, \texttt{Ox}, \texttt{Od} for \msvc, and \texttt{O0-O3} for GCC/ICC). 
Due to various constraints (\eg some speed testing programs do not support x86), we cannot compile all programs using all optimization flags.
In total, we have 588 compiled binaries for SPEC CPU2017.

The second dataset is SPEC CPU2006~\cite{spec2017cpu}, the former generation of SPEC benchmarks. It includes 19 C/C++/Fortran programs. 
We follow the same configurations as Andriesse~\etal\cite{andriesse2016depth} to compile SPEC CPU2006 so that we can easily obtain the ground truth for function and instruction boundaries. For example, we use the legacy GCC-5.1.1 to compile on Linux platforms. However, the Windows compilation and configuration are not available and no longer maintained after we have contacted the authors. Therefore, we use \msvc 2008 on a Windows XP virtual machine to compile the binaries for the Windows platform. 
To compensate for relatively small number of available binaries comparing to SPEC CPU2017, we turn on one extra optimization flag in GCC (\texttt{Os}), just to enlarge the pretraining dataset and introduce more byte patterns. 
As the default installations of GCC-5.1.1 and \msvc 2008 cannot compile certain legacy programs in SPEC CPU2006, we have in total 333 compiled binaries.

The third dataset is from BAP corpora~\cite{bao2014byteweight}. It includes 2,200 compiled binaries where 136 popular open-source programs (\eg \texttt{vim}) are on Windows platform, and the remaining 2,064 ELF binaries (from \texttt{coreutils}, \texttt{binutils}, and \texttt{findutils} packages) are compiled on Linux platform. Both sets are further divided equally into x86 and x64 binaries.

\begin{table*}[!t]
\footnotesize
\setlength{\tabcolsep}{12pt}
\centering
\renewcommand{\arraystretch}{1}

\caption{General statistics of all binaries and performance of \sys on all binaries (discussed in Section~\ref{subsec:result}). 
The byte sequences are of length 512 (see Appendix Section~\ref{sec:hyperparm}).
We use 10\% of the binaries for training (in finetuning) across all datasets and the corresponding train-test overlap rate is shown in each row.}
\label{tab:dataset}

\begin{tabular}{c|l|c|c|c|l|l|l|l}
\toprule
\multirow{2}{*}{Dataset} & \multirow{2}{*}{\begin{tabular}[c]{@{}c@{}}Total \# \\ Binaries\end{tabular}} & \multirow{2}{*}{Platform} & \multirow{2}{*}{Compiler} & \multirow{2}{*}{ISA} & \multicolumn{1}{c|}{\multirow{2}{*}{\# Binaries}} & \multicolumn{1}{c|}{\multirow{2}{*}{\# Bytes}} & \multicolumn{1}{c|}{\multirow{2}{*}{\begin{tabular}[c]{@{}c@{}}\# Byte\\ Sequences\end{tabular}}} & \multicolumn{1}{c}{\multirow{2}{*}{\begin{tabular}[c]{@{}c@{}}Train-test\\ Overlap\end{tabular}}} \\ 
 &  &  &  &  &  &  &  &  \\ \midrule 
\multirow{4}{*}{\begin{tabular}[c]{@{}c@{}}SPEC\\ 2017\end{tabular}} & \multirow{4}{*}{588} & \multirow{2}{*}{Linux} & \multirow{2}{*}{GCC-9.2} & x86 & 120 & 198,019,576 & 386,757 & 0.001\%  \\ \cline{5-9} 
 &  &  &  & x64 & 224 & 464,906,401 & 908,021 & 0.2\% \\ \cline{3-9} 
 &  & \multirow{2}{*}{Windows} & \multirow{2}{*}{\msvc-2019} & x86 & 88 & 175,057,814 & 341,910 & 0.93\% \\ \cline{5-9} 
 &  &  &  & x64 & 156 & 955,201,152 & 1,865,628 & 0.97\% \\ \hline
\multirow{4}{*}{\begin{tabular}[c]{@{}c@{}}SPEC\\ 2006\end{tabular}} & \multirow{4}{*}{333} & \multirow{2}{*}{Linux} & \multirow{2}{*}{GCC-5.1.1} & x86 & 90 & 55,637,428 & 108,667 & 0.002\% \\ \cline{5-9} 
 &  &  &  & x64 & 95 & 74,006,029 & 144,543 & 0\% \\ \cline{3-9} 
 &  & \multirow{2}{*}{Windows} & \multirow{2}{*}{\msvc-2019} & x86 & 76 & 40,417,016 & 78,940 & 0.36\% \\ \cline{5-9} 
 &  &  &  & x64 & 72 & 48,403,456 & 94,538 & 0.21\% \\ \hline
\multirow{4}{*}{BAP} & \multirow{4}{*}{2,200} & \multirow{2}{*}{Linux} & \multirow{2}{*}{\begin{tabular}[c]{@{}c@{}}GCC-4.7.2 \&\\ ICC-14.0.1\end{tabular}} & x86 & 1,032 & 138,547,936 & 270,602 & 1\% \\ \cline{5-9} 
 &  &  &  & x64 & 1,032 & 145,544,012 & 284,266 & 1.1\% \\ \cline{3-9} 
 &  & \multirow{2}{*}{Windows} & \multirow{2}{*}{\begin{tabular}[c]{@{}c@{}}\msvc-2010 \&\\ 2012 \& 2013\end{tabular}} & x86 & 68 & 29,093,888 & 56,824 & 0.4\% \\ \cline{5-9} 
 &  &  &  & x64 & 68 & 33,351,168 & 65,139 & 2.3\% \\ \bottomrule
\end{tabular}
\end{table*}

\noindent\textbf{Baselines.} We use IDA~Pro v7.4~\cite{hex2008ida}, Ghidra v9.1~\cite{ghidra}, and objdump as baselines.
We also compare \sys with the other two research prototypes that achieve state-of-the-art results on recovering function boundaries. The first one is Nucleus~\cite{andriesse2017compiler}, which is based on the control-flow analysis. The second one is from Shin~\etal\cite{shin2015recognizing}, who designed a bi-RNN~\cite{shin2015recognizing} to recover function boundaries. Since Shin~\etal did not release their source code, we re-implement bi-RNN in PyTorch following the same setup described in their paper. Specifically, we adopt the 2-layer architecture with 16 hidden layer size, which reportedly achieves the best result.

\noindent\textbf{Label collection.}
We collect the ground truth labels for function boundaries and assembly instructions using the debug symbols and source code of the binaries.
(1) To get function boundaries for Windows binaries, we parse PDB files using Dia2dump~\cite{dia2dump}.
For Linux binaries, we parse DWARF information using the pyelftools~\cite{pyefltools}.
Like Bao~\etal\cite{bao2014byteweight} and many disassemblers, we do not count thunks in Windows binaries and trampolines in the \texttt{.plt} section of Linux binaries as functions. We remove thunks and trampolines from the outputs of function recovery tools that count them as functions, to ensure that they are not inadvertently seen as false positives.
(2) To get assembly instructions, we collect instruction boundaries as our ground truth, rather than the instructions themselves. This allows us to modularize the design of our model. Given the instruction boundaries, we can then deterministically map the bytes in each boundary to their corresponding instruction. To collect the instruction boundaries, we use source-level information to guide a linear disassembler to get the labels for most bytes, and then manually analyze the remaining bytes~\cite{andriesse2016depth}. The linear disassembler we use is the Capstone library~\cite{quynh2014capstone}.

\noindent\textbf{Metrics.}
Both tasks (\ie recovering functions and instructions) have the \emph{imbalanced label} problem. For example, the number of function boundaries (both start and end) in the SPEC CPU2006 dataset accounts for only 1.05\% of the total number of bytes.
A predictor that always outputs ``not boundary'' can achieve a 98.95\% accuracy.
Therefore, we use alternative metrics, described in the following.

\emph{Precision, recall, F1 score.}
We use precision ($P$), recall ($R$), and $F1$ score to measure the actual performance of \sys and all other tools.
Consider recovering function boundaries. Let $TP$ (true positive) denote the number of correctly predicted (\eg function) boundaries, $FP$ denote the number of incorrectly predicted boundaries, $FN$ denote the number of incorrectly predicted not-boundaries, and $TN$ denote the number of correctly predicted non-boundaries. $P=TP/(TP+FP)$, $R=TP/(TP+FN)$, and $F1=2\cdot P\cdot R/(P+R)$.

\emph{Perplexity.}
We use perplexity (PPL) to evaluate the masked LM pretraining task, which intuitively measures how ``confused'' the model is about the masked bytes being equal to the ground truth bytes. 
The less confused is the model, the smaller the perplexity.
Specifically, $PPL=2^{-\frac{1}{N}\sum \log(p(x))}$. Here $p(x)$ is the probability produced by the pretrained model on the masked byte $x$ of being the ground-truth byte value. The summation applies to all the masked bytes. The smallest perplexity is thus 1 when the model is 100\% certain $p(x)=1$). 

\emph{Train-test overlap rate.}
We use the \emph{train-test overlap rate} of the dataset to quantify the introduced learning difficulties. The intuition is that it is easier for an ML model to achieve good testing performance if the testing samples also appear in the training set.
We define the train-test overlap rate as the percentage of overlapping byte sequences between testing and training. Particularly, we measure the percentage of test byte sequences that appear in the training set. This percentage reflects how challenging the test set is, as the model cannot just memorize all sequences for predicting the masked bytes~\cite{andriesse2017compiler}.

\noindent\textbf{Pretraining setup.}
To strictly separate the binaries used for pretraining, finetuning, and testing, we pretrain \sys on pairs of datasets shown in Table~\ref{tab:dataset}, \emph{and finetune on the third dataset}.
For example, all reported results of SPEC CPU2017 in Table~\ref{tab:overall_result} are finetuned (detailed in the following) on the model pretrained on SPEC CPU2006 and BAP.

Note that pretraining \emph{does not get access to any labeled data of any downstream task} (\eg disassembled instructions or function boundaries).
Therefore, the common practice in transfer learning is often to pretrain on large-scale corpora that can potentially include the finetuning data (without labels)~\cite{devlin2018bert, goyal2017accurate, rajpurkar2016squad, dagan2005pascal}.
However, in our evaluation, we strictly keep all pretraining, finetuning, and testing data separated to ensure that no testing binaries in finetuning are shared across the training dataset even though it is unfair to us. 

We pretrain models for each dataset pairs (described above) for 10 epochs. 
We hold out 4 binaries, each randomly selected from Windows x86, Windows x64, Linux x86, and Linux x64, respectively, as our validation set. We keep the pretrained model weights that achieve the best validation PPL and load its copy for finetuning experiment.

\noindent\textbf{Finetuning setup.}
For finetuning each subset of binaries (\ie in each row of Table~\ref{tab:dataset}), we randomly choose \emph{only 10\% of the dataset as our training set} and treat the remaining binaries as the testing set. As opposed to the typical train-test split, where the training set is often larger than the testing set, our setting creates a very challenging scenario for ML.
It increases the possibility that a large portion of test byte sequences and their underlying patterns do not appear in the training data.

As Andriesse \etal\cite{andriesse2017compiler} argue, one limitation of the BAP dataset is that duplicated functions are prevalent.
Consequently, high accuracy for recovering function boundaries (\eg bi-RNN~\cite{shin2015recognizing}) can be trivial to achieve on the BAP dataset. Our setting of using a small fraction of data as training mitigates this issue. 
For example, in Table~\ref{tab:dataset}, the highest train-test overlap rate is not more than 3\%.
Besides, on other datasets (SPEC CPU2017 and CPU2006) we have evaluated, the duplicated functions are much less (we do not statically link the libraries when compiling the program, and SPEC programs are collected from diverse domains).

We run 30 epochs for both finetuning \sys and training bi-RNN. 
We observe that \sys converges after 5 epochs of finetuning, but we selected 30 epochs for a fair comparison with bi-RNN, which often takes around 20 epochs to converge.
As Shin~\etal\cite{shin2015recognizing} trains their model for 2 hours on a CPU, another reason we choose 30 epochs to train bi-RNN is that we find 30 training epochs consumes at least 2 hours on our CPUs for all datasets.

\section{Evaluation}
\label{sec:eval}

Our evaluation aims to answer the following questions.
\begin{itemize}
    \item RQ1: How accurate is \sys in recovering function boundaries and instructions compared to other tools?
    \item RQ2: How robust is \sys under different platforms, compilers, architectures, and optimizations, compared to other tools?
    \item RQ3: How fast is \sys compared to other tools?
    \item RQ4: How efficient is \sys in terms of saving labeling effort and training epochs compared to other tools?
    \item RQ5: How effective is pretraining, and how does it help finetuning tasks?
\end{itemize}

\subsection{RQ1: Accuracy}
\label{subsec:result}

We first evaluate how accurate \sys is for the tasks of recovering function boundaries and assembly instructions.

\noindent\textbf{Overall result.}
Table~\ref{tab:overall_result} shows that, on average, \sys achieves an F1 score of 99\% on recovering function boundaries, 17.2\% better than the second-best tool,
and 99.7\% on recovering assembly instructions, outperforming all other tools across all platforms, ISAs, compilers, datasets, and number of training and testing binaries.

\begin{table*}[!t]
\footnotesize
\setlength{\tabcolsep}{7.5pt}
\centering
\renewcommand{\arraystretch}{1}

\caption{Results on recovering function boundaries averaged over all compiler optimization levels. For results on recovering instructions, we compute on the highest optimization levels considering lower optimization levels are too simple for all tools.}.
\label{tab:overall_result}

\begin{tabular}{c|c|c||ccccc|ccccc}
\toprule
\multirow{2}{*}{Dataset} & \multirow{2}{*}{Platform} & \multirow{2}{*}{ISA} & \multicolumn{5}{c|}{Recovering Function Boundaries F1 (\%)} & \multicolumn{5}{c}{Recovering Instructions F1 (\%)} \\
 &  &  & \sys & Nucleus & bi-RNN & IDA & Ghidra & \sys & bi-RNN & IDA & Ghidra & objdump \\ \midrule 
\multirow{4}{*}{\begin{tabular}[c]{@{}c@{}}SPEC\\ 2017\end{tabular}} & \multirow{2}{*}{Linux} & x86 & \textbf{98.4} & 55.4 & 79.9 & 91.8 & 89.0 & 99.9 & 87.1 & 95.9 & 94.6 & \textbf{100.0}$^{\dagger}$ \\ \cline{3-13} 
 &  & x64 & \textbf{99.1} & 55.0 & 79.2 & 90.2 & 89.5 & 99.9 & 88.9 & 95.8 & 95.9 & \textbf{100.0}$^{\dagger}$ \\ \cline{2-13} 
 & \multirow{2}{*}{Windows} & x86 & \textbf{99.1} & 60.8 & 73.8 & 67.6 & 70.4 & 99.2 & 82.3 & 96.7 & 92.1 & \textbf{99.3} \\ \cline{3-13} 
 &  & x64 & \textbf{98.9} & 65.0 & 78.4 & 78.0 & 71.6 & \textbf{99.4} & 81.9 & 97.1 & 93.1 & 99.3 \\ \hline
\multirow{4}{*}{\begin{tabular}[c]{@{}c@{}}SPEC\\ 2006\end{tabular}} & \multirow{2}{*}{Linux} & x86 & \textbf{98.2} & 57.2 & 86.7 & 95.7 & 92.2 & 99.9 & 89.0 & 96.3 & 95.5 & \textbf{100.0}$^{\dagger}$ \\ \cline{3-13} 
 &  & x64 & \textbf{98.7} & 56.8 & 73.8 & 92.8 & 92.0 & 99.8 & 85.9 & 96.4 & 94.9 & \textbf{100.0}$^{\dagger}$ \\ \cline{2-13} 
 & \multirow{2}{*}{Windows} & x86 & \textbf{99.4} & 68.2 & 78.5 & 77.9 & 76.3 & \textbf{99.7} & 89.9 & 98.1 & 94.5 & 99.1 \\ \cline{3-13} 
 &  & x64 & \textbf{98.3} & 56.8 & 72.7 & 90.1 & 86.2 & \textbf{99.4} & 86.2 & 97.9 & 95.7 & 99.4 \\ \hline
\multirow{4}{*}{BAP} & \multirow{2}{*}{Linux} & x86 & \textbf{99.5} & 61.5 & 74.1 & 59.0 & 57.2 & N/A$^*$ & N/A$^*$ & N/A$^*$ & N/A$^*$ & N/A$^*$ \\ \cline{3-13} 
 &  & x64  & \textbf{98.7} & 53.5 & 79.0 & 58.3 & 56.5 & N/A$^*$ & N/A$^*$ & N/A$^*$ & N/A$^*$ & N/A$^*$ \\ \cline{2-13} 
 & \multirow{2}{*}{Windows} & x86 & \textbf{99.5} & 69.0 & 80.1 & 89.9 & 87.0 & N/A$^*$ & N/A$^*$ & N/A$^*$ & N/A$^*$ & N/A$^*$ \\ \cline{3-13} 
 &  & x64 & \textbf{99.4} & 70.0 & 81.4 & 90.5 & 80.6 & N/A$^*$ & N/A$^*$ & N/A$^*$ & N/A$^*$ & N/A$^*$ \\ \hline
 \multicolumn{3}{c||}{Average} & \textbf{99.0} & 60.8 & 78.1 & 81.8 & 79.0 & \textbf{99.7} & 86.4 & 96.8 & 94.4 & 99.6 \\ \bottomrule
\multicolumn{13}{l}{\scriptsize $^*$The BAP corpus does not contain source code and PDB files, which are necessary to obtain the assembly instruction ground truth.}\\
\multicolumn{13}{l}{\scriptsize $^{\dagger}$GCC does not generate inline data, so a simple linear disassembler can achieve 100\% F1 score~\cite{andriesse2016depth}. objdump always fails to identify inline data, but because} \\ 
\multicolumn{13}{l}{\scriptsize inline data makes up only a tiny fraction (\textless 1\%) of the code section, objdump's overall F1 score is high.} \\
\end{tabular}
\end{table*}


We note that the linear disassembler objdump achieves a high F1 score at recovering instructions. Even though it might seem counter-intuitive, the high F1 score results from the fact that objdump naively disassembles all bytes as code, making no attempt to identify inline data. Since inline data only account for a tiny fraction of bytes (usually $<$1\%), simply disassembling every byte as instruction is sufficient to achieve a high F1 score~\cite{andriesse2016depth}. objdump achieves a perfect F1 score on binaries compiled by GCC, because GCC does not generate inline data. 
However, objdump cannot handle the hard cases, but achieves a high F1 score because the hard cases are rare. 
To handle inline data, the recursive traversal disassemblers IDA~Pro and Ghidra act more conservatively, which sacrifice their accuracy in easier cases and result in lower F1 scores. In contrast, \sys can both identify inline data, such as jump tables (see Section~\ref{sec:case} for details), and maintain a high F1 score.



In Figure~\ref{fig:overall}, we show in further detail that \sys outperforms other tools by even greater margin at recovering function boundaries, when tested on high optimization levels (\texttt{O3} for GCC and ICC; \texttt{O2} for \msvc).
While Shin~\etal\cite{shin2015recognizing} (bi-RNN) reports \textgreater95\% F1 score in recovering function boundaries on all metrics (precision, recall, and F1) on the BAP dataset, we find that its performance is not as ideal in this setting (high optimization levels, small training set and low train-test overlap rate, and inclusion of other datasets such as SPEC CPU2017).

We noticed that in our experiments Nucleus has a lower F1 score compared to its performance in \cite{andriesse2017compiler}, and worked with the Nucleus team to investigate potential causes. With their help, we found 3 types of function patterns prevalent in samples in our datasets that Nucleus struggled to detect: functions ending with tailjumps; functions with instructions after a return, that are only reachable indirectly; and functions reached only via jumps. These cases often cause compounding errors, where failing to detect one function causes the tool to fail to detect related functions. The findings are consistent with the analysis of Nucleus error cases in \cite{andriesse2017compiler}. We concur with the authors' opinion that in some of these cases Nucleus's predictions are usable results for human analysts, despite differing from the ground truth generated from symbolic information.

The lower F1 score is potentially caused by compiler changes since the Nucleus tool was written, which increased the use of the above function patterns, and the inclusion of ICC, a compiler on which Nucleus was not previously tested.
The Nucleus team is working to analyze the issue more completely.



\begin{figure*}[!t]
\centering



\includegraphics[scale=.75]{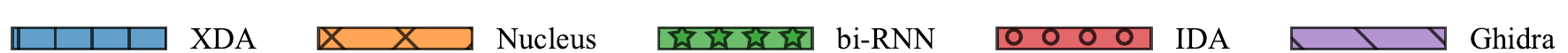}

\subfloat[Function - Linux x86]{
\includegraphics[width=0.22\linewidth]{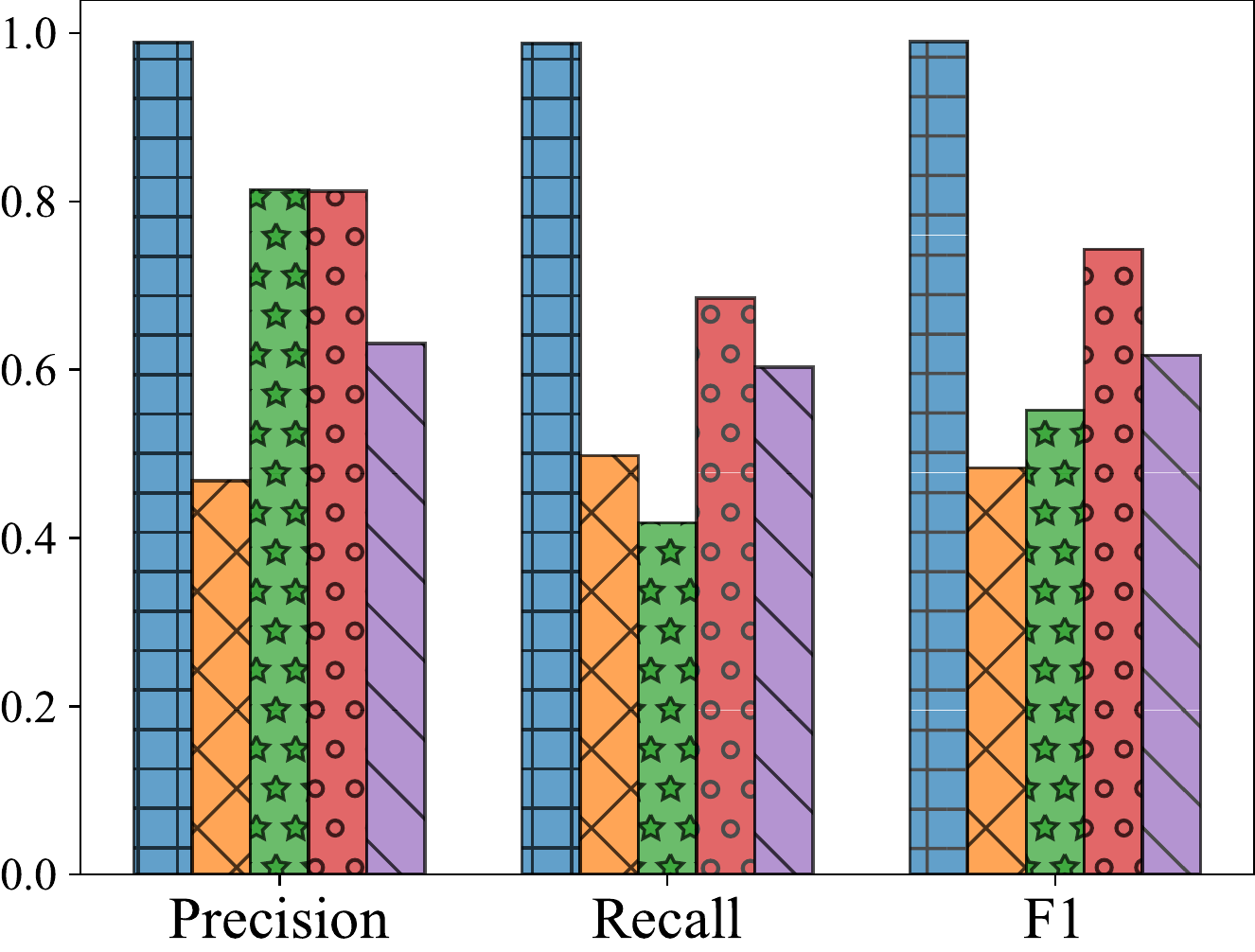}
\label{subfig:func_linux_x86}}
\subfloat[Function - Linux x64]{
\includegraphics[width=0.22\linewidth]{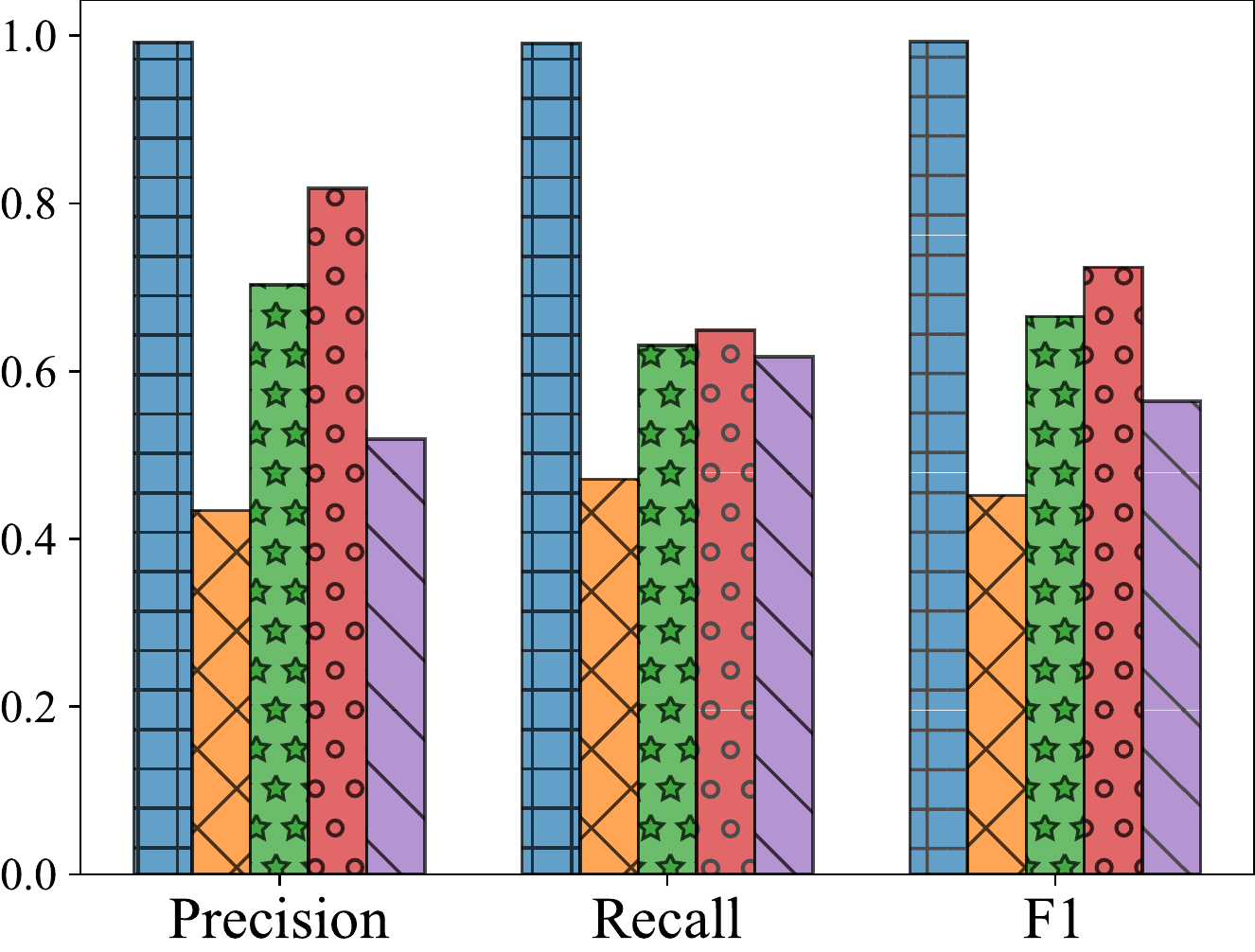}
\label{subfig:func_linux_x64}}
\subfloat[Function - Windows x86]{
\includegraphics[width=0.22\linewidth]{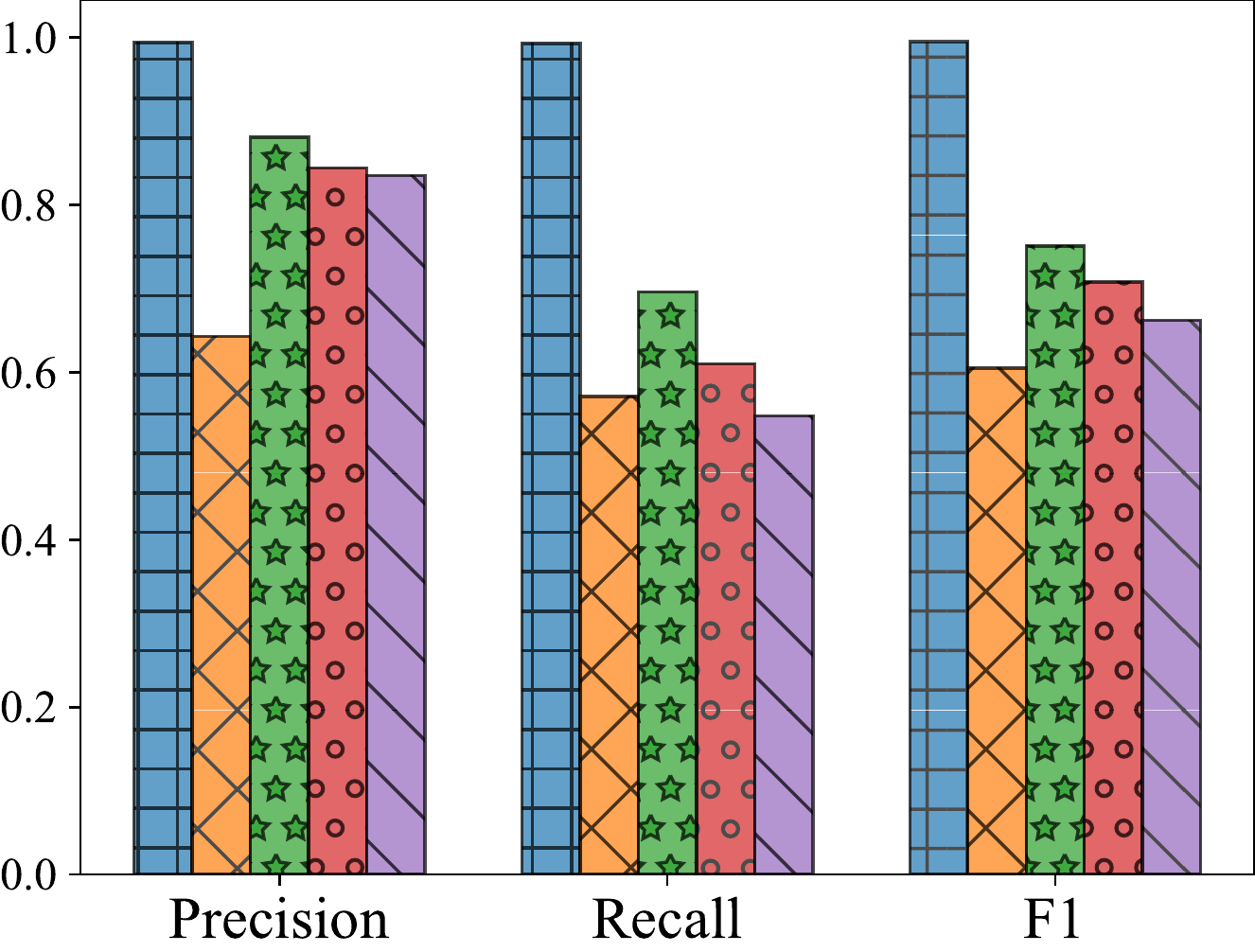}
\label{subfig:func_win_x86}}
\subfloat[Function - Windows x64]{
\includegraphics[width=0.22\linewidth]{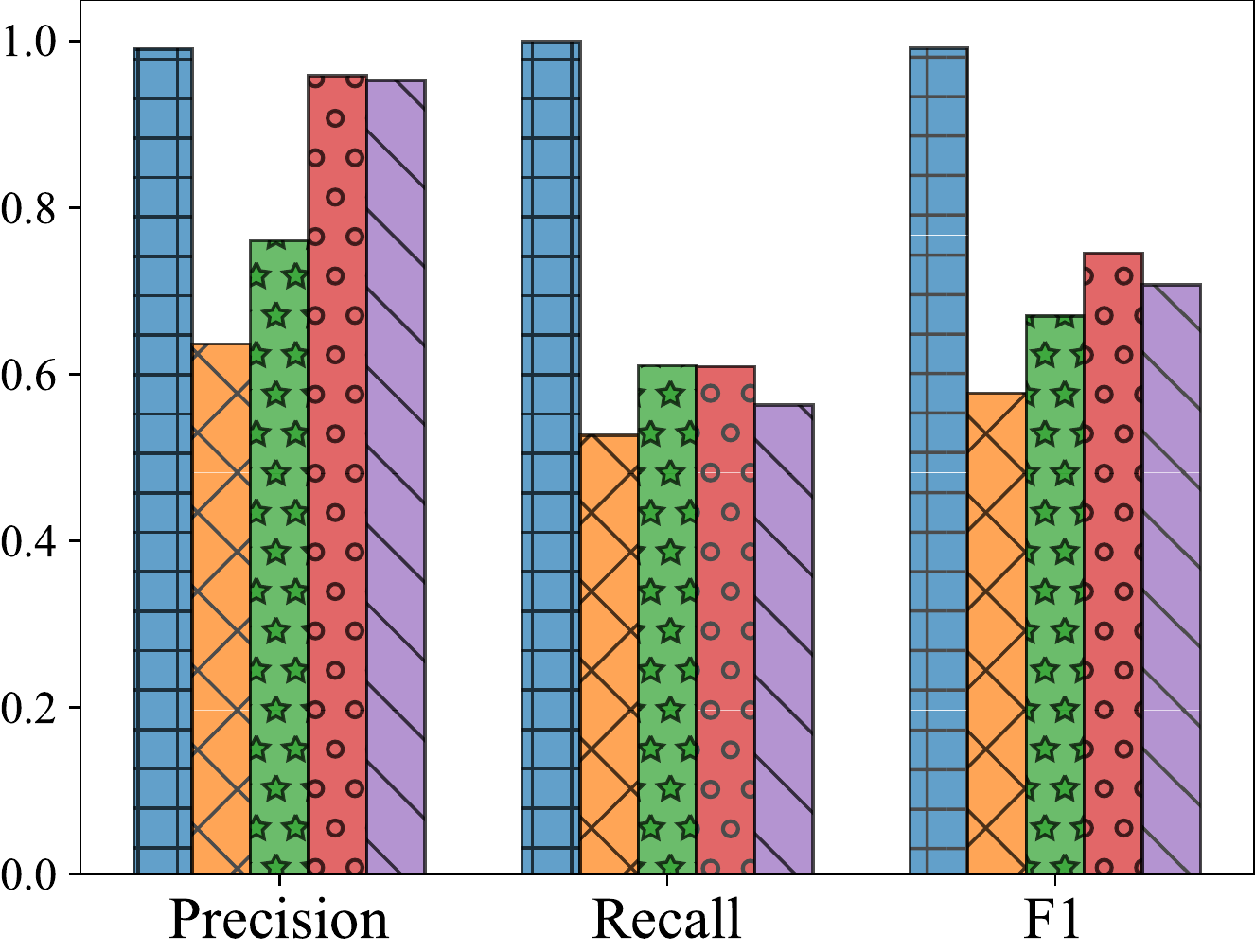}
\label{subfig:func_win_x64}}

\caption{Precision, recall, and F1 score of recovering function boundaries by \sys and other tools on four types of binaries categorized by the platforms and ISAs, on the highest compiler optimization levels.}
\label{fig:overall}
\end{figure*}

\noindent\textbf{Generalizability.}
We test how \sys generalizes to unseen byte sequences. Specifically, we vary the train-test overlap rate and compare the test F1 score achieved by \sys and bi-RNN.
We use SPEC CPU2017 binaries compiled on Windows x64 with optimization level \texttt{Ox} as our target for this test.
As described in Section~\ref{sec:impl}, we finetune on the model that has been pretrained on SPEC CPU2016 and BAP corpus.
We choose 4 target train-test overlap rate (20\%, 40\%, 60\%, and 80\%).
For each rate, we select as the training set random sequences from the entire testing sequences to construct the target train-test overlap rate.

\begin{table}[!t]

\footnotesize
\setlength{\tabcolsep}{14pt}
\centering
\renewcommand{\arraystretch}{1}

\caption{F1 score (\%) of \sys and bi-RNN on recovering function boundaries on varying train-test overlap rate.}
\label{tab:robust_overlap}

\begin{tabular}{c|cccc}
\toprule
 & \multicolumn{4}{c}{Train-test Overlap Rate} \\
 & 20\% & 40\% & 60\% & 80\% \\ \midrule 
bi-RNN & 70.1 & 82.3 & 89.8 & 96.5 \\
\sys & 99.1 & 99.5 & 99.8 & 99.9 \\ \bottomrule
\end{tabular}
\end{table}

Table~\ref{tab:robust_overlap} shows the testing F1 scores of both \sys and bi-RNN with different train-test overlap rate. 
Note that in Table~\ref{tab:overall_result}, \sys performs reasonably well on low train-test overlap rate (\textless 3\%, see Table~\ref{tab:dataset}). When we increased the train-test overlap rate, we found that \sys still reaches a 99\%+ F1 score after finetuning for 30 epochs.
However, bi-RNN obtains a high F1 score only at a high train-test overlap rate (\textgreater 80\%).
This supports findings by Andriesse~\etal\cite{andriesse2017compiler} that the performance of previous ML-based tools is biased.
One possible explanation is that bi-RNN relies heavily on memorizing syntactic function boundary patterns, and it fails to generalize when the patterns are absent in the training set.

\noindent\textbf{Generalizability to real-world software projects.}
Besides the dataset in Table~\ref{tab:dataset}, we also test the strict generalizability of \sys on real-world datasets.
In total, we collected 10 real-world popular opensource real-world software projects, including Curl-7.71.1, Diffutils-3.7, GMP-6.2.0, ImageMagick-7.0.10, Libmicrohttpd-0.9.71, LibTomCrypt-1.18.2, OpenSSL-1.0.1f and OpenSSL-1.0.1u, PuTTy-0.74, SQLite-3.34.0, and Zlib-1.2.11.
Note that we neither pretrain nor finetune \sys on these software projects, but just use them to test how \sys performs on the unseen binary programs.
We thus compile them using GCC-7.5 on Linux x64 with 4 optimizations (\texttt{O0}-\texttt{O3}).

\begin{table}[!t]
\footnotesize
\setlength{\tabcolsep}{10pt}
\centering
\renewcommand{\arraystretch}{1}

\caption{F1 score of \sys's function boundary recovery (pretrained on BAP and SPEC CPU2006 and finetuned on SPEC CPU2017 x64 binaries compiled on Linux with GCC) on unseen binaries collected from popular opensource projects.}
\label{tab:more_data}

\begin{tabular}{r|l|l|l|l}
\toprule
\textbf{}     & \texttt{O0} & \texttt{O1} & \texttt{O2} & \texttt{O3} \\ \midrule
Curl          & 98.6        & 98.5        & 98.6        & 98.2        \\ \hline
Diffutils     & 98.7        & 98.7        & 98.8        & 98.6        \\ \hline
GMP           & 99.2        & 98.8        & 98.9        & 98.6        \\ \hline
ImageMagick   & 98.4        & 98.3        & 98.2        & 98.2        \\ \hline
Libmicrohttpd & 98.9        & 98.8        & 98.9        & 98.7        \\ \hline
LibTomCrypt   & 99.0        & 99.0        & 98.7        & 98.6        \\ \hline
OpenSSL       & 98.4        & 98.3        & 98.4        & 98.2        \\ \hline
PuTTy         & 98.3        & 98.3        & 98.2        & 98.1        \\ \hline
SQLite        & 98.8        & 98.7        & 98.4        & 98.3        \\ \hline
Zlib          & 98.9        & 98.9        & 99.0        & 98.8        \\ \bottomrule
\end{tabular}
\end{table}

Table~\ref{tab:more_data} shows the testing F1 score of each software project and optimization achieved by the model pretrained on BAP and SPEC CPU2006, and finetuned (in 30 epochs) on SPEC CPU2017 x64 binaries compiled on Linux.
We find that \sys achieves at least 98.1 (and up to 99.1) F1 score at recovering function boundaries even when none of these binaries are seen during pretraining and finetuning, which is close to the average performance that \sys obtains in Table~\ref{tab:overall_result}.

\noindent\textbf{Tuning false positives/negatives.}
Some downstream binary analysis tasks, which potentially use \sys as the building block, are highly sensitive to false positives and negatives, depending on their nature.
For example, reverse engineering malware can be intolerant of false negatives (\eg the malicious code is overlooked), while binary re-writer cannot afford false positives (\eg the data treated as code and getting re-written corrupts the entire program). 

While \sys achieves 99\%+ F1 score, with only a minimal number of false positives and negatives (Figure~\ref{fig:overall}), we discuss a possible mechanism to tune the false positives/negatives to improve its practicality for diverse downstream tasks.
A common practice to control the tradeoff between false positive/negative is to threshold the model’s output probability. 
For example, we can predict a certain byte as a function boundary only if the model’s output probability on that byte is greater than a threshold. The model can thus be tuned to act more conservatively/aggressively in predicting function boundaries by varying the threshold.

Specifically, we use the ROC (Receiver Operating Characteristics) curve and its AUC (Area Under the Curve) score to quantify the trade-off between false positive/negative. 
Intuitively, ROC curve measures under different threshold (\ie beyond what confidence score does the model predict as boundary), what is the model's resulting false postive rate and false positive rate.
The higher the AUC score of the ROC curve, the better the model.
Therefore, given such curve, To tune the false positives, we can (ideally) choose the threshold that achieves 0\% false positive rate with 100\% true positive rate.

\begin{figure}[!t]
\centering

\includegraphics[width=0.75\linewidth]{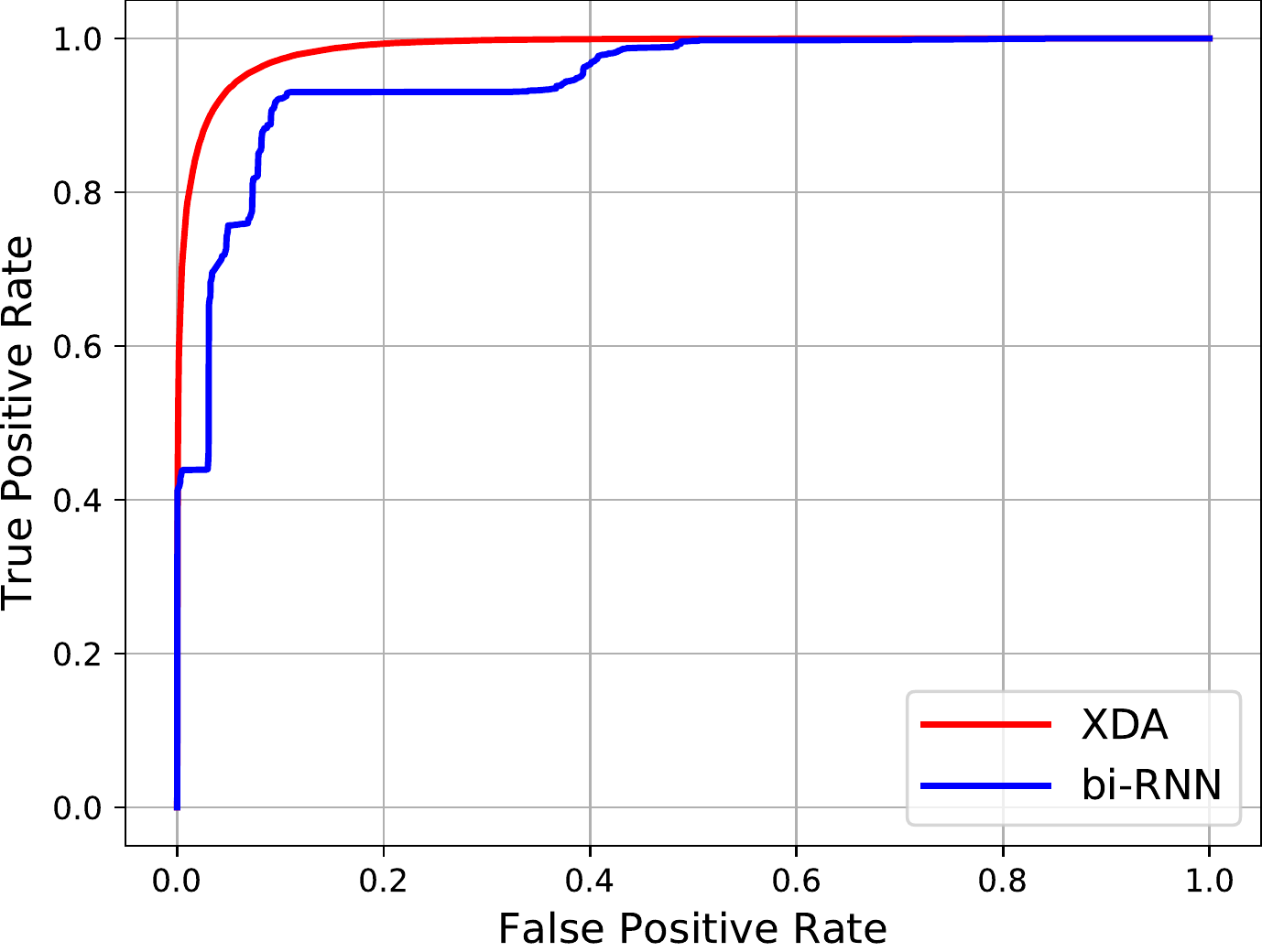}

\caption{The ROC curve of \sys and bi-RNN, when finetuning and testing on SPEC CPU2017 Windows x64 binaries compiled by \msvc.}
\label{fig:roc}
\end{figure}

Figure~\ref{fig:roc} shows the ROC curves of \sys and bi-RNN, when they are finetuned for 30 epochs and tested for recovering function boundaries on SPEC CPU2017 Windows x64 binaries compiled on Windows by \msvc. 
We find that \sys has higher AUC score (0.994) than that of bi-RNN (0.9), which indicates that \sys consistently outperforms bi-RNN on different thresholds.

\subsection{RQ2: Robustness}
\label{subsec:robust_analysis}
In this section, we analyze the robustness of \sys to high compiler optimization.
In Section~\ref{subsec:result}, we can see that \sys generalizes well across different platforms, compilers (different platforms imply different compilers), and ISAs, but we have not explicitly shown the robustness on different optimization levels.
For example, in Figure~\ref{fig:overall}, we can only see \sys is robust to binaries compiled with the \emph{highest optimization}.
However, we have not seen whether \sys and other disassemblers can remain robust under \emph{different} optimizations.
Therefore, we evaluate the robustness of \sys under varying optimization levels and compare them with bi-RNN.

\noindent\textbf{Robustness to different optimizations.} 
We deliberately chose SPEC CPU2017 x64 binaries compiled on the Windows platform as our target for this test, because these binaries represent the hardest cases in our dataset for an ML model to learn. 
First, it has less overlapping functions between training and testing than BAP, so simple patterns seen in the training data may not be useful~\cite{andriesse2017compiler}.
Second, the function length in SPEC CPU2017 is often very diverse, as opposed to the BAP dataset that has mostly similar-length common library functions.
Third, we observe that \msvc uses inline data in the code section for jump tables, while modern versions of GCC keep the code and data separated~\cite{andriesse2016depth}.
Such features make it even harder for recovering function boundaries, as inline data can disrupt the patterns in the code.
Finally, the total number of bytes in SPEC CPU2017 is large (see Table~\ref{tab:dataset}). Since we only keep 10\% for training, the large testing set makes it more challenging to obtain good results.
Similar to previous setting, the model to be finetuned is pretrained on SPEC CPU2016 and BAP corpus.

\begin{figure}[!t]
\centering

\includegraphics[scale=.9]{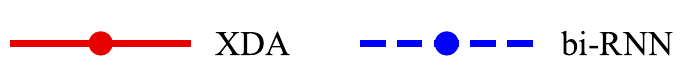}

\subfloat[\texttt{O1}]{
\includegraphics[width=0.47\linewidth]{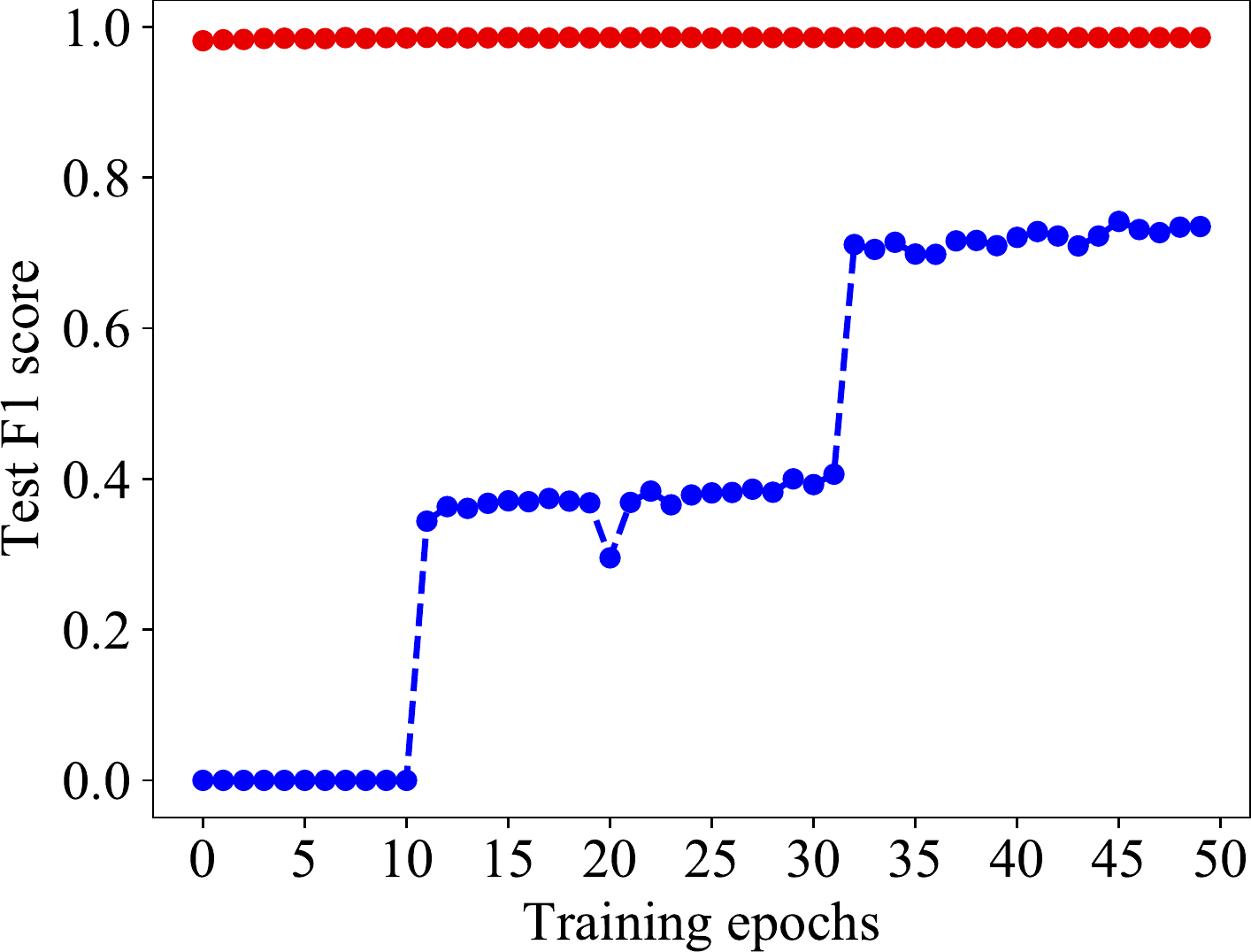}
\label{subfig:O1}}
\subfloat[\texttt{O2}]{
\includegraphics[width=0.47\linewidth]{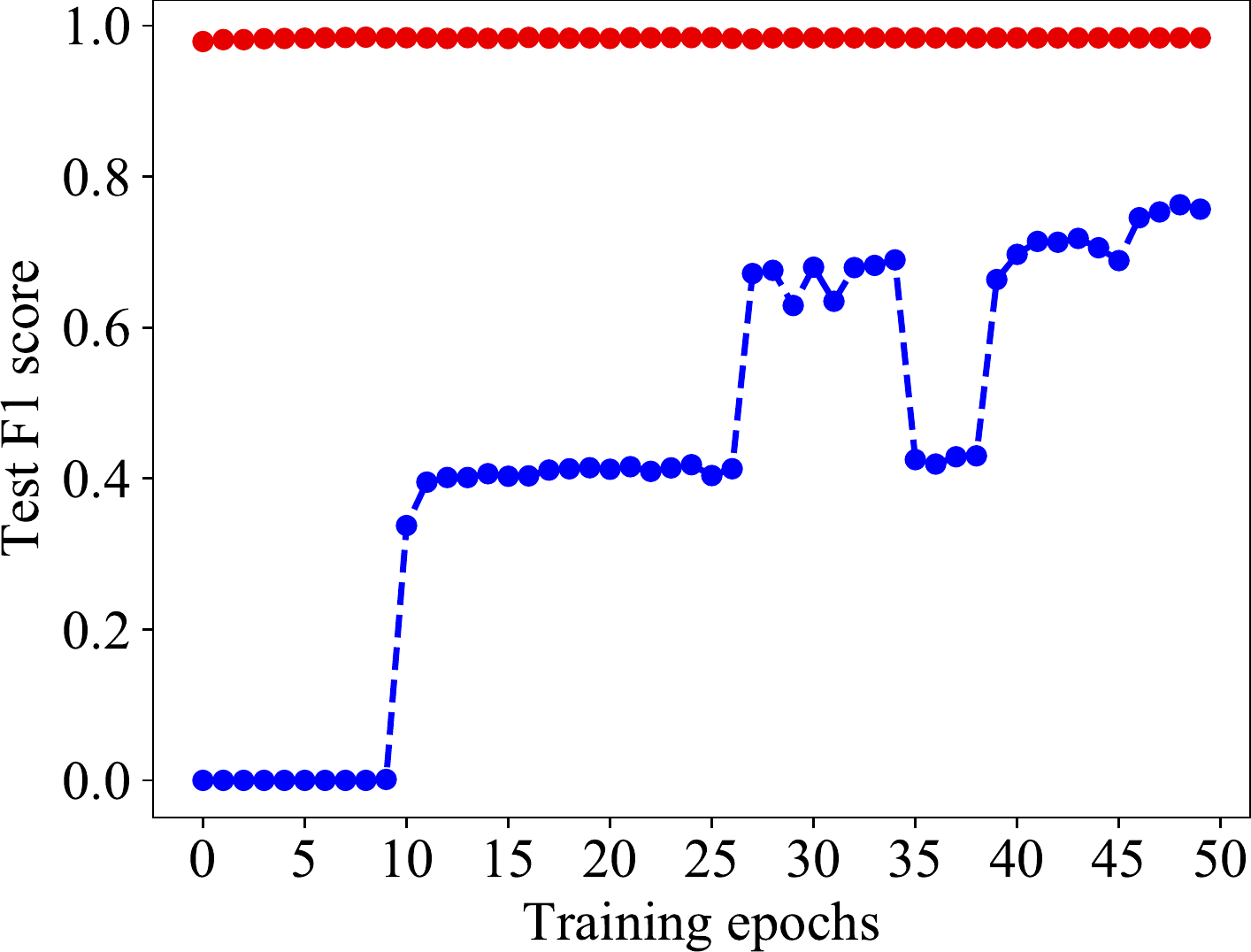}
\label{subfig:O2}}

\subfloat[\texttt{Od}]{
\includegraphics[width=0.47\linewidth]{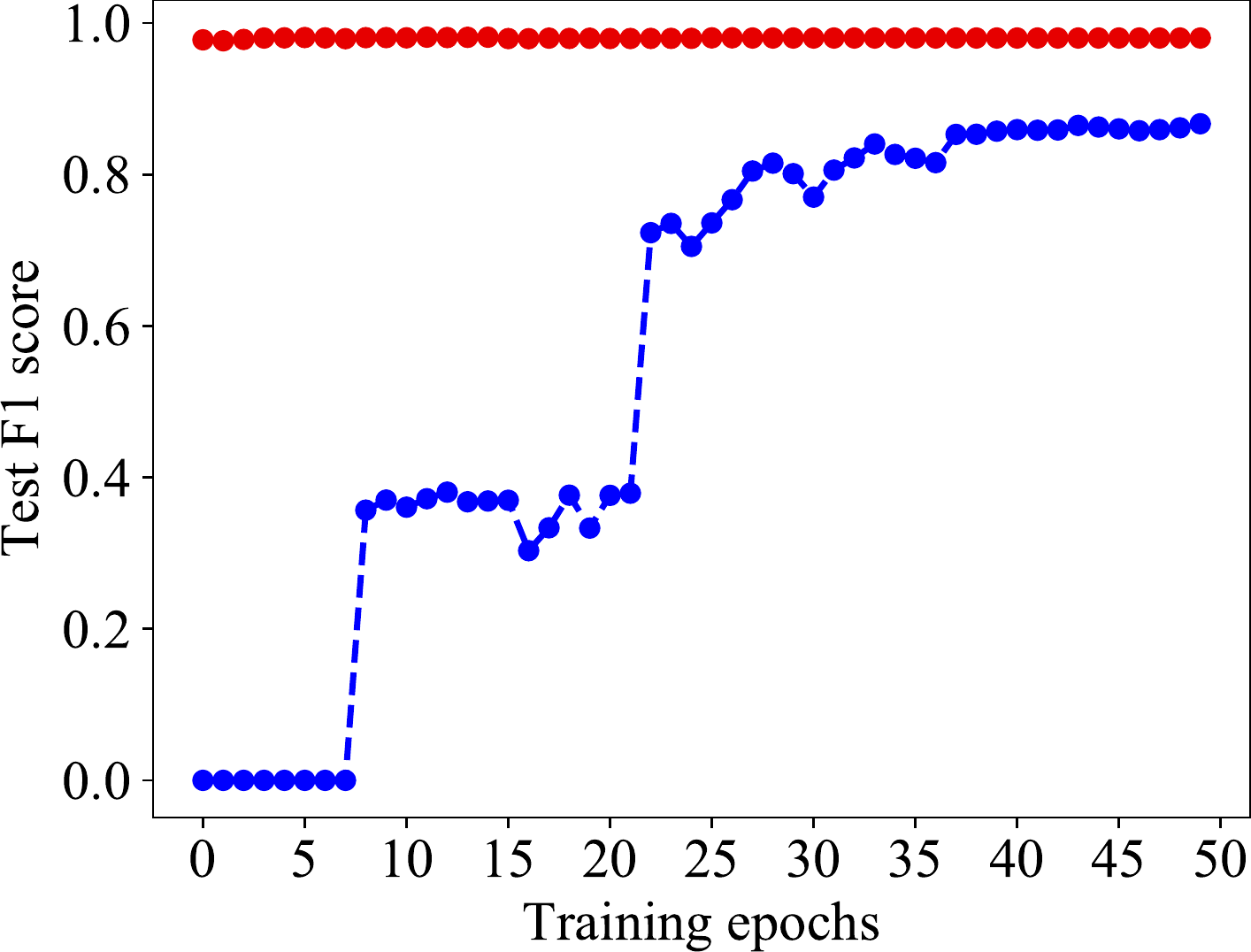}
\label{subfig:Od}}
\subfloat[\texttt{Ox}]{
\includegraphics[width=0.47\linewidth]{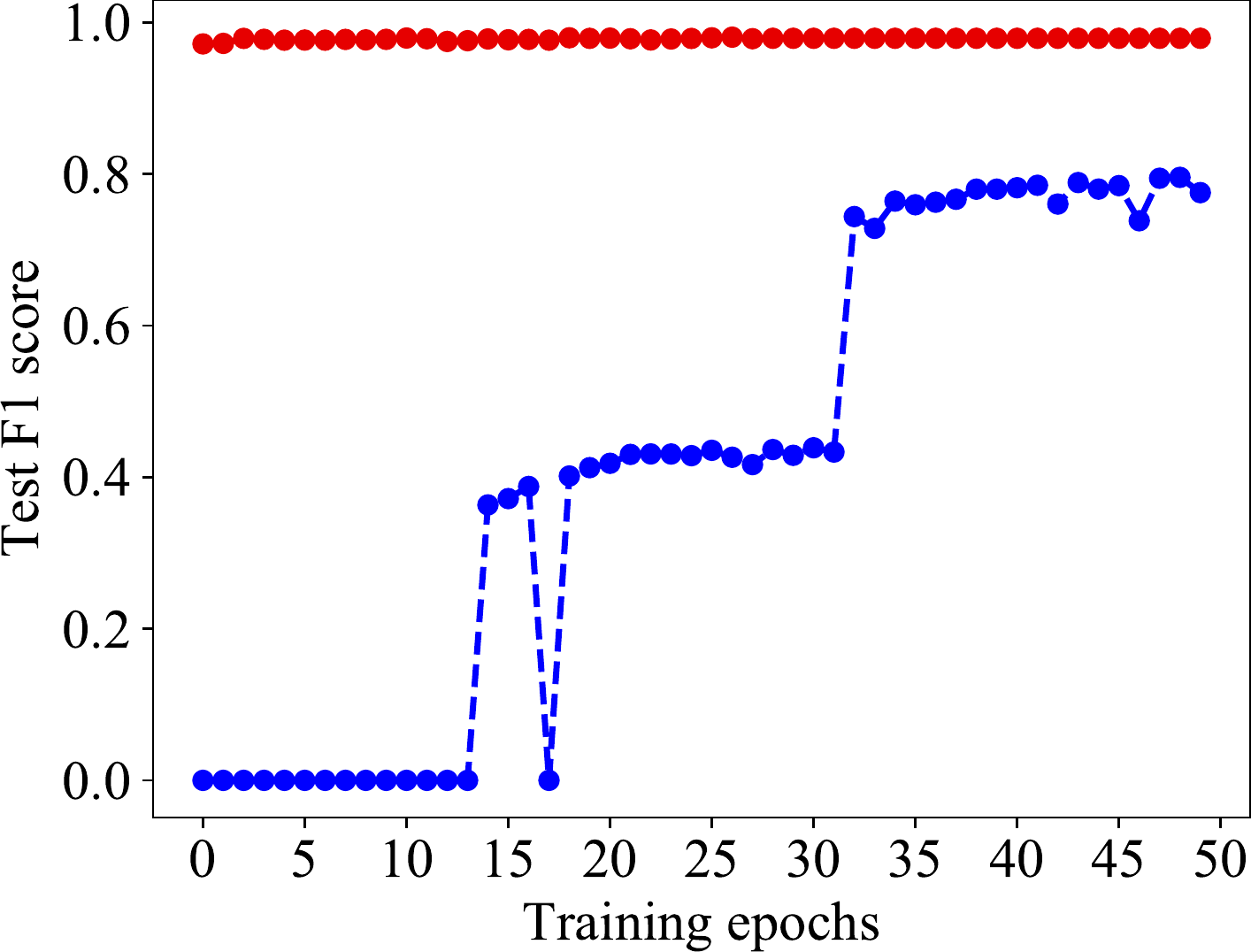}
\label{subfig:Ox}}

\caption{Comparison between \sys and bi-RNN on recovering function boundaries with varying optimization levels (\texttt{O1, O2, Ox, Od}).}
\label{fig:robust_opt}
\end{figure}

We separate the binaries based on their optimization flags (\texttt{O1, O2, Od, Ox}) and evaluate \sys's performance of recovering function boundaries.
Figure~\ref{fig:robust_opt} shows the testing F1 scores in 50 training epochs of both \sys and bi-RNN under each optimization level.
We find that \sys's performance remains robust under different optimization levels (\textgreater98.8\% F1), while bi-RNN performs worse on binaries with higher optimization levels and remains at least 20\% worse than \sys on all optimization levels.
In addition, impressively, \sys can always reach a very high F1 score even after the first epoch, while bi-RNN struggles in the first 30 epochs.

\noindent\textbf{Transferability across optimization levels.}
The above experiment tests \sys on different optimization levels, but it still finetunes/trains on all optimization levels.
In this experiment, we test if finetuning \sys on \emph{only one optimization level} can still achieve a good F1 score on another optimization level. We use the same dataset (SPEC CPU2017) and same pretrained model (on SPEC CPU2006 and BAP) described above.
We use bi-RNN as the baseline.

Table~\ref{tab:transfer} shows that \sys always achieves a 98.5\%+ F1 score when finetuning on binaries with one optimization level and testing on binaries with another optimization level. 
However, bi-RNN has an apparent decrease in F1 scores when the training and testing binaries are compiled in different optimization levels.
The excellent transferability of \sys implies that the model learns machine code semantics that is robust across different optimization levels. 

\begin{table}[!t]
\footnotesize
\setlength{\tabcolsep}{6pt}
\centering
\renewcommand{\arraystretch}{1}

\caption{Test F1 score (\%) of \sys and bi-RNN trained and tested on different optimization flags.}
\label{tab:transfer}

\begin{tabular}{c|c|c|c|c|c}
\toprule
 & \backslashbox{Train OPT}{Test OPT} & \texttt{O1} & \texttt{O2} & \texttt{Od} & \texttt{Ox} \\ \midrule 
\multirow{4}{*}{bi-RNN} & \texttt{O1} & \textbf{81} & 80 & 47 & 2.8 \\ \cline{2-6}
 & \texttt{O2} & 44 & \textbf{85} & 81 & 75 \\ \cline{2-6}
 & \texttt{Od} & 34.5 & 4.1 & \textbf{85.2} & 43.6 \\ \cline{2-6}
 & \texttt{Ox} & 80 & 39 & 44 & \textbf{87} \\ \midrule 
\multirow{4}{*}{\sys} & \texttt{O1} & \textbf{99.8} & 98.6 & 98.8 & 98.5 \\ \cline{2-6}
 & \texttt{O2} & 99.8 & \textbf{98.7} & 99 & 98.9 \\ \cline{2-6}
 & \texttt{Od} & 99.6 & 98.5 & \textbf{99} & 98.7 \\ \cline{2-6}
 & \texttt{Ox} & 99.7 & 98.7 & 98.8 & \textbf{98.9} \\ \bottomrule
\end{tabular}
\end{table}

\noindent\textbf{Robustness to obfuscated binaries.}
We also checked how \sys performs on obfuscated binaries, even when we do not pretrain nor finetune \sys on any obfuscated binaries.
We obfuscate all our collected real-world software projects (described in Section~\ref{subsec:result}) using 5 types of obfuscations (OBF) by Hikari~\cite{hikari} on x64 -- an obfuscator based on \texttt{clang-8}.
The obfuscation strategies include bogus control flow (\texttt{bcf}), control flow flatening (\texttt{cff}), register-based indirect branching (\texttt{ibr}), basic block splitting (\texttt{spl}), and instruction substitution (\texttt{sub}).
We turn off the compiler optimization in case it optimizes away the obfuscated code.
We then evaluate the robustness of \sys on each of the software project shown in Table~\ref{tab:more_data} obfuscated by each of the obfuscation types.

Table~\ref{tab:obf} lists the F1 score achieved by \sys at recovering function boundaries, with the same pretraining and finetuning setup described in Section~\ref{subsec:result}. We observe that \sys remains robust for all obfuscated software projects, achieving at least 98 and up to 99.2 F1 score.
Notably, \sys performs the best on the binaries obfuscated by instruction substitution (\texttt{sub}), which is intuitive as such obfuscation only tampers with the arithmetic operation~\cite{hikari} without affecting the function prologues and epilogues
Again, we neither pretrained nor finetuned \sys on any obfuscated binaries.
Indeed, in practice, we can easily collect a large number of obfuscated binaries for pretraining and finetuning.
We thus expect that the model can have even stronger results when it can learn the dependencies between bytes in the obfuscated binaries.

\begin{table}[!t]
\footnotesize
\setlength{\tabcolsep}{8pt}
\centering
\renewcommand{\arraystretch}{1}

\caption{F1 score of \sys's function boundary recovery on unseen binaries obfuscated with 5 different obfuscation types.}
\label{tab:obf}

\begin{tabular}{r|l|l|l|l|l}
\toprule
\textbf{}     & \texttt{bcf} & \texttt{cff} & \texttt{ibr} & \texttt{spl} & \texttt{sub} \\ \midrule
Curl          & 98.3         & 98.5         & 98.6         & 98.5         & 99.0         \\ \hline
Diffutils     & 98.6         & 98.7         & 98.4         & 98.7         & 99.1         \\ \hline
GMP           & 99.0         & 98.9         & 98.9         & 98.5         & 99.2         \\ \hline
ImageMagick   & 98.3         & 98.3         & 98.1         & 98.0         & 98.4         \\ \hline
Libmicrohttpd & 98.6         & 98.7         & 98.8         & 98.6         & 98.9         \\ \hline
LibTomCrypt   & 98.7         & 98.6         & 98.7         & 98.6         & 98.9         \\ \hline
OpenSSL       & 98.3         & 98.9         & 98.6         & 98.9         & 99.0         \\ \hline
PuTTy         & 98.2         & 98.1         & 98.1         & 98.0         & 98.3         \\ \hline
SQLite        & 98.7         & 98.6         & 98.1         & 98.4         & 98.8         \\ \hline
Zlib          & 98.0         & 98.4         & 98.1         & 98.6         & 99.0         \\ \bottomrule
\end{tabular}
\end{table}

\subsection{RQ3: Execution Time}
\label{subsec:runtime}

\noindent\textbf{Comparison with other tools.} We compare the speed of recovering function boundaries between \sys and two well-known manual-written disassemblers, IDA~Pro and Ghidra. Specifically, we choose 4 x64 binaries with different sizes in SPEC CPU2017 compiled by \msvc with \texttt{O2}. We run both IDA and Ghidra in command-line mode to eliminate the runtime overhead of their GUIs.
Table~\ref{tab:runtime} shows the time taken by each tool to recover the function boundaries from a binary. \sys outperforms both IDA and Ghidra by up to 38$\times$.

\noindent\textbf{GPU vs. CPU.}
We compare the speed of \sys running on CPUs versus on GPUs. Since \sys's architecture (\ie self-attention layers) can be significantly accelerated by GPUs, its inference time on CPU is up to 10$\times$ slower than on a GPU.
However, as seen in Table~\ref{tab:runtime}, \sys on CPU still typically outperforms IDA and Ghidra by up to 3$\times$. 

\begin{table}[!t]

\footnotesize
\setlength{\tabcolsep}{3pt}
\centering
\renewcommand{\arraystretch}{1}

\caption{Execution time of \sys, IDA, and Ghidra (in seconds) on binaries with different size. We also show the speedup achieved by \sys over the second-fastest tool.}
\label{tab:runtime}

\begin{tabular}{lr|cccc|c}
\toprule
\multirow{3}{*}{Binary} & \multirow{3}{*}{Size} & \multicolumn{4}{c|}{Speed} & \multirow{3}{*}{\begin{tabular}[c]{@{}c@{}}\sys \\ speedup\end{tabular}} \\
 &  & \sys & \sys & \multirow{2}{*}{IDA} & \multirow{2}{*}{Ghidra} \\ 
 &  & GPU & CPU &  &  &  \\ \midrule 
\texttt{specrand\_is} & 556K & \textbf{1.6} & 12 & 4.8 & 16.4 & 3$\times$ \\
\texttt{omnetpp\_r} & 4.1M & \textbf{3.5} & 29 & 92.3 & 146.3 &  26$\times$ \\
\texttt{cpuxalan\_r} & 7.8M & \textbf{4.9} & 50.1 & 192.0 & 185.7 & 38$\times$ \\
\texttt{blender\_r} & 22.0M & \textbf{16.1} & 136.0 & 317.1 & 246.2 & 15$\times$ \\ \bottomrule
\end{tabular}
\end{table}

\subsection{RQ4: Training Efficiency}
\label{subsec:effort}

In this section, we aim to quantify the training efficiency of \sys in comparison with other ML-based tools.
Intuitively, since pretraining has already encoded rich knowledge useful for downstream tasks, \sys only requires a small amount of the labeled data and training epochs for finetuning to achieve good results on downstream tasks.
In particular, we compare \sys with bi-RNN at recovering function boundaries, on (1) the number of labeled training data required to achieve certain F1 scores, and (2) the number of training epochs needed to achieve specific F1 scores, using the same training data.
We use Windows x64 SPEC CPU2017 binaries compiled by \msvc with all 4 optimizations as our dataset. We base the finetuning on the pretrained model on SPEC CPU2006 and BAP.

\noindent\textbf{Labeled training data.}
We test if \sys requires less training data than bi-RNN to achieve a comparable F1 score.
To control the number of labeled training data, we sort the binaries by their size and start with training on the largest binary. Then we gradually increase the number of training binaries (\eg the second largest, and so on) until the testing F1 score surpasses a chosen threshold.
We choose the last half of the sorted binaries as the test set. We choose five F1 score thresholds (0.9, 0.7, 0.5, 0.3, 0.1), and test how many training binaries are required for both \sys and bi-RNN to go beyond these chosen thresholds.
We train both models for 30 epochs.

\begin{table}[!t]

\footnotesize
\setlength{\tabcolsep}{6pt}
\centering
\renewcommand{\arraystretch}{1}

\caption{The \emph{required number of training binaries and epochs} by \sys and bi-RNN to surpass the F1 score thresholds.}
\label{tab:required_binaries_epochs}

\begin{tabular}{c|c|ccccc}
\toprule
 & & \multicolumn{5}{c}{F1 score threshold} \\
 & & $>0.9$ & $>0.7$ & $>0.5$ & $>0.3$ & $>0.1$ \\ \midrule
\multirow{2}{*}{\begin{tabular}[c]{@{}c@{}}\# training\\ binaries\end{tabular}} & bi-RNN & N/A & 8 & 4 & 1 & 1 \\ \cline{2-7}
 & \sys & \textbf{2} & \textbf{1} & \textbf{1} & \textbf{1} & \textbf{1} \\ \midrule
\multirow{2}{*}{\begin{tabular}[c]{@{}c@{}}\# training\\ epochs\end{tabular}} & bi-RNN & N/A & 28 & 24 & 14 & 13 \\ \cline{2-7}
 & \sys & \textbf{2} & \textbf{2} & \textbf{1} & \textbf{1} & \textbf{1} \\ \bottomrule
\end{tabular}
\end{table}

The upper two rows of Table~\ref{tab:required_binaries_epochs} shows that \sys always needs only 1 training binary to surpass 0.7 testing F1 score while 2 binaries to go above 0.9. However, bi-RNN needs at least 8 training binaries to obtain better than 0.7 F1 score. Moreover, it cannot go beyond 0.9, even if it uses up all the remaining 78 training binaries (total 156 binaries except half used for testing).

\noindent\textbf{Training epochs.}
Now we test if \sys can be trained in less training epochs than bi-RNN to achieve comparable F1 score.
Following the above setup, we keep only the largest binary as the training file and take the last half of the sorted binaries as the testing set. 
We then show the number of required training epochs F1 score in Table~\ref{tab:required_binaries_epochs} (lower two rows) to surpass the same chosen testing threshold defined above.
We observe that the number of training epochs required by \sys is much less than bi-RNN. For example, \sys needs at most 2 training epochs to surpass all F1 thresholds, while bi-RNN takes 28 epochs to go beyond 0.7.

\subsection{RQ5: Pretraining Effectiveness}
\label{subsec:pretrain}

\noindent\textbf{Pretraining robustness.}
We first evaluate the robustness of \sys's pretraining task, where we test if \sys could transfer the learned knowledge on one platform/compiler/architecture to another for the masked LM task. 
The rationale is that the compiler idiom varies significantly between platforms and architectures (\eg calling conventions), generating completely different binaries. 
Therefore, if the pretrained model on \emph{all} binaries can generalize to different compiler configurations and instruction sets, we can use a single pretrained model for all downstream tasks instead of pretraining on a specific dataset every time for each finetuning task.


Note that in previous experiments, we pretrained multiple models with different dataset partitions to strictly separate the pretraining and finetuning data (see Section~\ref{sec:impl}) \emph{just for the fair comparison with other baselines}. 
However, in practice, we can always collect a large corpus of binaries in the wild for pretraining, and always reuse the same pretrained model for finetuning on different labeled datasets. Pretraining can thus be a one-time cost.\footnote{We study how results in Table~\ref{tab:overall_result} can be further improved if we pretrain on all available datasets in Appendix Section~\ref{subsec:pretrain_all}.}

We combine all datasets in Table~\ref{tab:dataset}, and pretrain 5 different models where their training data is partitioned based on the platforms/compilers or ISAs.
Specifically, we pretrain models on (1) \emph{all} binaries, (2) all \emph{Linux x86} binaries, (3) all \emph{Linux x64} binaries, (4) all \emph{Windows x86} binaries, and (5) all \emph{Windows x64} binaries.
We leave 4 programs from SPEC CPU2017 dataset out as the test binaries, each from one platform (Windows and Linux) and one ISA (x86 and x64).
To test pretrained models on \emph{all} binaries, we combine all 4 programs as the test set. 

\begin{table}[!t]
\footnotesize
\setlength{\tabcolsep}{3pt}
\centering
\renewcommand{\arraystretch}{1}

\caption{Test PPL of pretrained models. We also include cross-dataset PPL (\eg train on x86 and test on x64).}
\label{tab:pretrain}

\begin{tabular}{c|c|c|c|c|c}
\toprule
\backslashbox{Training}{Testing} & All & \begin{tabular}[c]{@{}c@{}}Linux x86\\ GCC/ICC\end{tabular} & \begin{tabular}[c]{@{}c@{}}Linux x64\\ GCC/ICC\end{tabular} & \begin{tabular}[c]{@{}c@{}}Win x86\\\msvc\end{tabular} & \begin{tabular}[c]{@{}c@{}}Win x64 \\ \msvc\end{tabular} \\ \midrule 
All & \textbf{1.41} & 1.56 & 1.25 & 1.45 & 1.26 \\ \hline
\multirow{2}{*}{\begin{tabular}[c]{@{}c@{}}Linux x86\\ GCC/ICC\end{tabular}} & \multirow{2}{*}{2.26} & \multirow{2}{*}{\textbf{1.55}} & \multirow{2}{*}{2.31} & \multirow{2}{*}{2.14} & \multirow{2}{*}{2.88} \\ 
 &  &  &  &  & \\ \hline
\multirow{2}{*}{\begin{tabular}[c]{@{}c@{}}Linux x64\\ GCC/ICC\end{tabular}} & \multirow{2}{*}{2.41} & \multirow{2}{*}{2.57} & \multirow{2}{*}{\textbf{1.25}} & \multirow{2}{*}{3.03} & \multirow{2}{*}{2.25} \\
 &  &  &  &  & \\ \hline
\multirow{2}{*}{\begin{tabular}[c]{@{}c@{}}Win x86\\ \msvc\end{tabular}} & \multirow{2}{*}{2.26} & \multirow{2}{*}{2.26} & \multirow{2}{*}{2.5} & \multirow{2}{*}{\textbf{1.44}} & \multirow{2}{*}{2.7} \\ 
 &  &  &  &  & \\ \hline
\multirow{2}{*}{\begin{tabular}[c]{@{}c@{}}Win x64\\ \msvc\end{tabular}} & \multirow{2}{*}{1.98} & \multirow{2}{*}{2.72} & \multirow{2}{*}{2.02} & \multirow{2}{*}{2.5} & \multirow{2}{*}{\textbf{1.26}} \\ 
 &  &  &  &  & \\ \bottomrule
\end{tabular}
\end{table}

Table~\ref{tab:pretrain} shows the test PPL \sys achieves after 10 epochs of pretraining. When pretrained on all binaries (first row), its PPL drops to at least 1.56.\footnote{The best PPL so far on natural language is around 3.6~\cite{liu2019roberta}}
We also test the generalizability of the pretrained model across different platforms and ISAs. 
We note that \sys also performs reasonably well across binaries. For example, PPL can drop to 2.5 when training on Windows x86 but testing on Linux x64, which is far below that of random guessing (\eg $2^{-\log(1/256)}=256$).
More importantly, \emph{pretraining on all binaries} (Table~\ref{tab:pretrain} first row) has obtained the best PPL to all platforms and ISAs (only worse than the cases where training and testing are from the same platforms and ISAs), because its training data includes all type of binaries.


\noindent\textbf{Benefit of pretraining.}
We quantify how much pretraining can improve the finetuning tasks.
Specifically, we test performance of recovering function boundaries on SPEC CPU2017 Windows x64 binaries compiled by \msvc (using 10\% random binary as the training set). 
We compare \sys's F1 scores achieved by varying pretraining data size. We compare pretrained model on (1) 100\% of pretraining data (\ie on SPEC CPU2006 and BAP), (2) 75\% of pretraining data, (3) 50\% of pretraining data, (4) 25\% of pretraining data, and (5) without pretraining. 
We also include the F1 scores of bi-RNN for comparison.

Figure~\ref{fig:ablation} shows that pretraining helps greatly in improving the F1 score of recovering function boundaries.
The model always reaches \textgreater94\% F1 score within the first epoch, \eg even when the model is pretrained on only 25\% of pretraining data.
With more epochs of finetuning, they can always reach \textgreater98\% F1 scores.
Besides, the more pretraining data, the better the finetuning performance.
Without pretraining, \sys's F1 scores gradually converges to 90\% after 28 epochs and are lower than bi-RNN in first 14 epochs.
One possible reason is that the underlying model of \sys has many trainable parameters (see Section~\ref{subsec:pretrain_method} and Appendix Section~\ref{sec:hyperparm}), which requires more training epochs and data than bi-RNN.

\begin{figure}[!t]
\centering

\includegraphics[width=.8\linewidth]{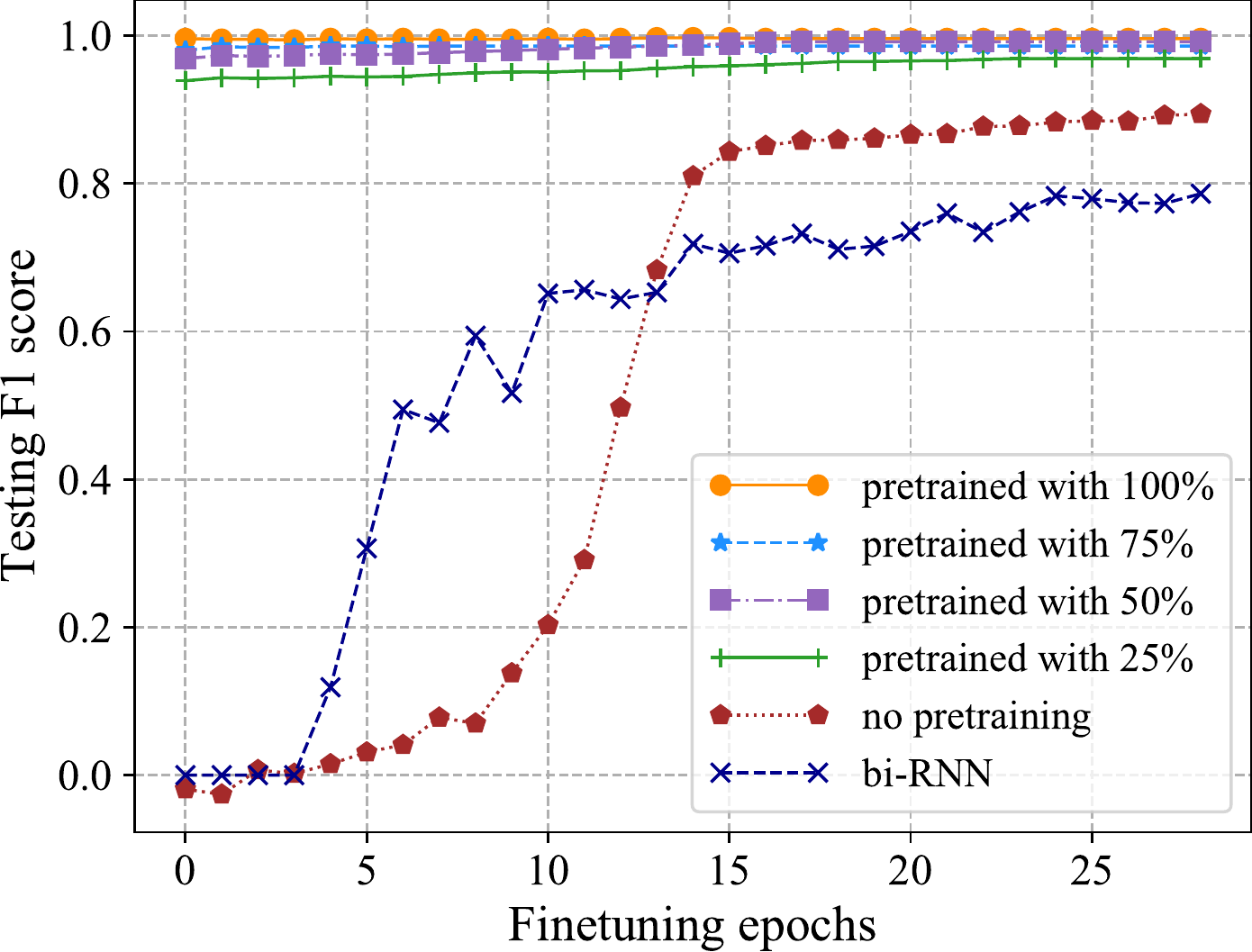}

\caption{Comparison of testing F1 score on recovering function boundaries between (1) with 100\% pretraining data, (2) with 66\% pretraining data, (3) with 33\% pretraining data, (4) without pretraining, and (5) bi-RNN.}
\label{fig:ablation}
\end{figure}

\noindent\textbf{Pretraining other neural architectures.}
In theory, any neural net architecture can be pretrained, including our bi-RNN baseline.
In this paper, we specifically pretrained on the Transformer architecture, as its self-attention layers are known to be amenable to parallelization using GPUs, supporting scalable pretraining.
In comparison, recurrent neural network, such as bi-RNN, is known for its limited scalability (due to its sequential nature) on longer sequences and larger datasets~\cite{vaswani2017attention}.
Therefore, given the scale of our pretraining dataset (over 5 gigabytes), training bi-RNN is prohibitively expensive (\eg our tests show bi-RNN is 100$\times$ slower than the Transformer training on the same amount of data), so we omit pretraining bi-RNN in this paper.

\section{Case Studies}
\label{sec:case}

We show some concrete examples to demonstrate that \sys learns various byte dependencies and semantics. We also illustrate the attentions of \sys when making predictions to help explain \sys's decision process.

\subsection{Probing Learned Semantics}
\label{subsec:probe_case}

\noindent\textbf{Predicting instructions.}
Consider the byte sequence shown in Figure~\ref{fig:predicting_instructions}, taken from the SPEC CPU2017 \texttt{sgcc} compiled by \msvc x64 with \texttt{O2}.
We mask out the bytes of whole instruction at \texttt{0x3f} and feed the sequence to \sys to predict the masked bytes.
We find that \sys can recover the correct bytes with high confidence (\ie \textgreater 70\%). We make following two interesting observations. These observations further justifies the usefulness of pretraining in helping recovering assembly instructions and function boundaries.

\begin{figure}[!t]
\centering

\includegraphics[width=0.99\linewidth]{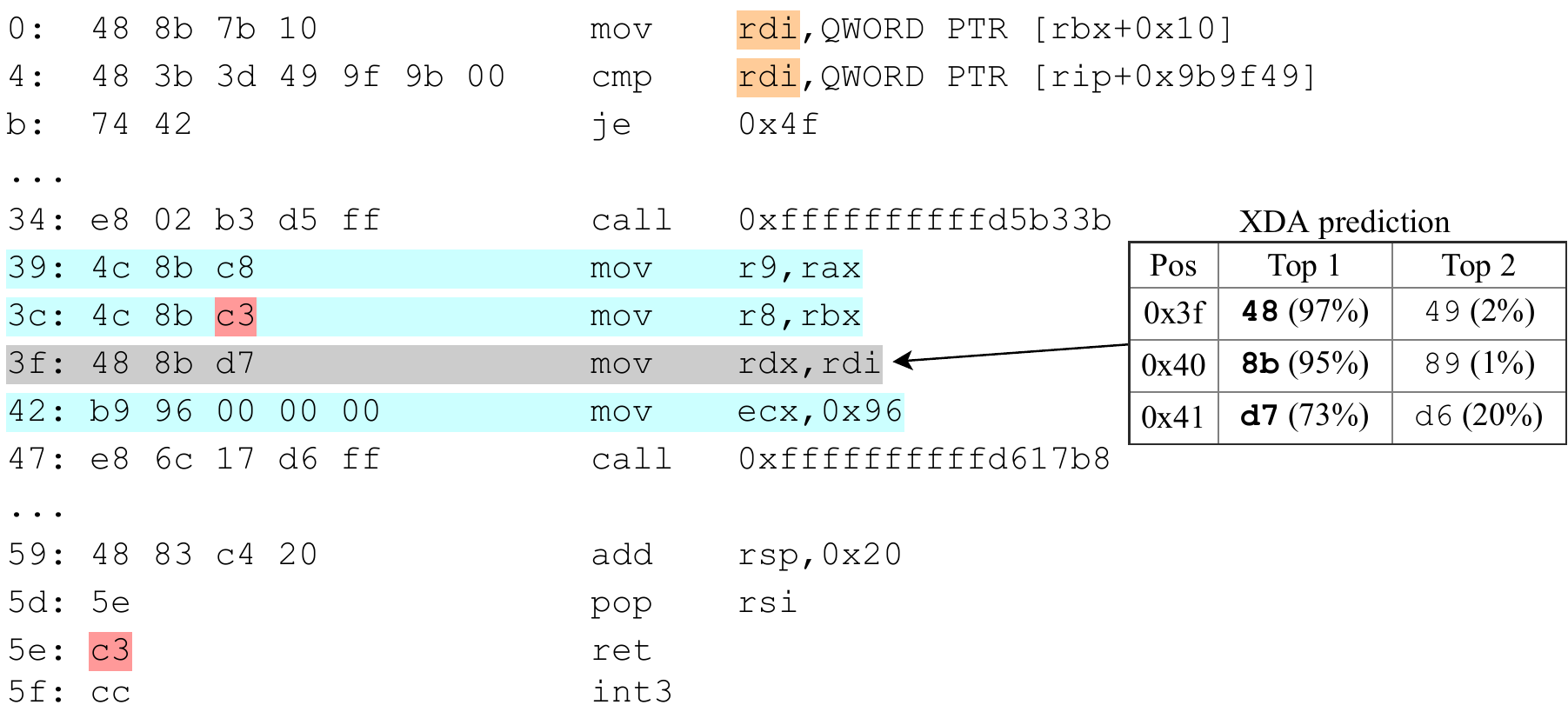}

\caption{We mask \colorbox{lightgray}{\texttt{48 8b d7}}, one of the \colorbox{lightblue}{argument-passing instructions} of the function call at \texttt{0x47}. \sys's top-2 predictions with confidence are shown on the right-hand side. We highlight byte \colorbox{lightred}{\texttt{c3}}, as predicting the masked bytes requires distinguishing the meaning of \texttt{c3}, which can be both \texttt{ret} or a register depending on the context. We highlight register \colorbox{lightorange}{\texttt{rdi}} where \sys likely leverages to predict \texttt{rdi} in the masked bytes.}
\label{fig:predicting_instructions}
\end{figure}

\emph{\sys learns calling convention.} As the masked instruction belong to the argument passing procedures, correctly predicting the exact instructions implies that \sys \emph{understands the calling conventions} of Windows x64 (\ie saving arguments to the registers in the order of \texttt{r9 r8 rdx rcx}~\cite{ms64call}). We can see that \sys is very confident that the first two bytes are \texttt{48 8b} (which translates to \texttt{mov rdx,-}), as it has seen the \texttt{r9 r8} appeared before and \texttt{ecx} (lower 32-bit of \texttt{rcx}). In contrast, it is less certain on the third byte, as moving which register's value to \texttt{rdx} is harder to infer. Therefore, we see 73\% confidence of \texttt{d7}, which will be decoded as \texttt{rdi}, while 20\% confidence of \texttt{d6}, which will be decoded as \texttt{rsi}.

\emph{\sys learns instruction semantics.} Note that the highlighted \texttt{c3} in the byte sequence have completely different meanings depending on the context. Recall that the input to \sys is only the plain byte sequence. As \texttt{c3} itself denote \texttt{ret} instruction, If \sys mistakenly treat \texttt{c3} at location \texttt{0x3d} as \texttt{ret}, the next most likely bytes would be a sequence of padding bytes \texttt{cc} that usually lie between functions for alignment. 
Therefore, the only explanation is \sys implicitly learns that \texttt{c3} at location \texttt{0x3d} is part of another instruction.

\noindent\textbf{Predicting the gaps between functions.}
Consider the same binary (SPEC CPU2017 \texttt{sgcc} program compiled by \msvc x64 with \texttt{O2}), where we feed a different byte sequence, as shown in Figure~\ref{fig:predicting_padding}, to \sys. We mask out all the padding bytes \texttt{cc} between two functions. The byte before the first \texttt{cc} is a function end at \texttt{0x5e}, and byte after the last \texttt{cc} is the function start at \texttt{0x84}.

\begin{figure}[!t]
\centering

\includegraphics[width=0.99\linewidth]{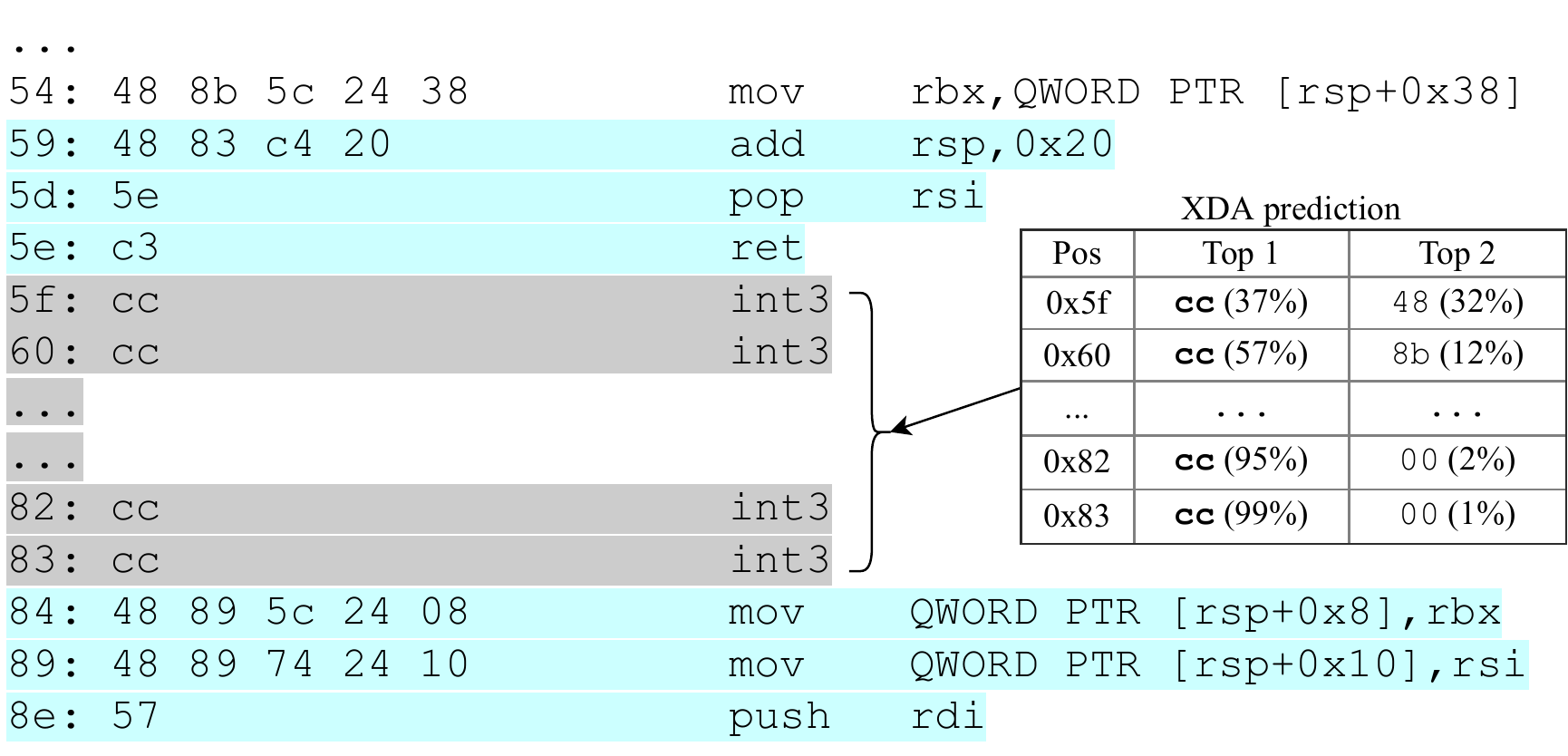}

\caption{We mask a sequence of padding bytes \colorbox{lightgray}{\texttt{cc}}, which reside between two functions. We highlight the typical function \colorbox{lightblue}{prologue/epilogue} that \sys leverages to make predictions for the masked bytes.}
\label{fig:predicting_padding}
\end{figure}

We can see that \sys predicts all the masked bytes correctly in the right-hand table.
As the model cannot leverage any patterns of padding between functions (we have masked out all \texttt{cc}), the only context left for \sys to correctly predict the masked bytes are the function epilogue at \texttt{0x5d} (\texttt{pop rsi}) to \texttt{0x5e} (\texttt{ret}), and the following likely function prologue at \texttt{0x84} (\texttt{mov QWORD PTR [rsp+0x8],rbx}) to \texttt{0x8e} (\texttt{push rdi}).
As \sys only sees byte sequences, we speculate that \sys must understand that the corresponding bytes before/after the masked bytes represent the function epilogue/prologue, in order to predict those masked bytes are padding.
We thus perform a sanity check to ask \sys to predict the function boundary. This time we load the \emph{finetuned} \sys and feed the same byte sequence without masking and ask the finetuned model to predict the function boundaries.
\sys then predicts that there is a function end at \texttt{0x5f} and a function start at \texttt{0x84}, which match the ground truth. We thus conclude that the task of predicting masked bytes can effectively help the model to recover function boundaries.

\noindent\textbf{Predicting jump table entries.}
Consider the \texttt{vim} compiled by Visual Studio x64 with \texttt{O1} in Figure~\ref{fig:predicting_jump_table_entries}, where we mask out one jump table entry. 
We can see that \sys correctly predicts all the masked bytes in the jump table with high confidence.
More interestingly, we find that while \sys can predict the least significant byte of the jump target (the left-most byte in the jump table entry), it is less confident than predicting other bytes. As the least significant byte is the finest-grained byte the determines the address of the jump target, it is understandable that it is extremely difficult to guess. Moreover, \sys predicts \texttt{8b} as the second possible candidate byte after \texttt{77}, as \texttt{8b} appears twice in the context in other jump table entries. This indicates \sys still leverage the pattern as an important hint, but it is not dominated by only pattern matching because it still predicts \texttt{77} as the most probable byte.

\begin{figure}[!t]
\centering

\includegraphics[width=0.99\linewidth]{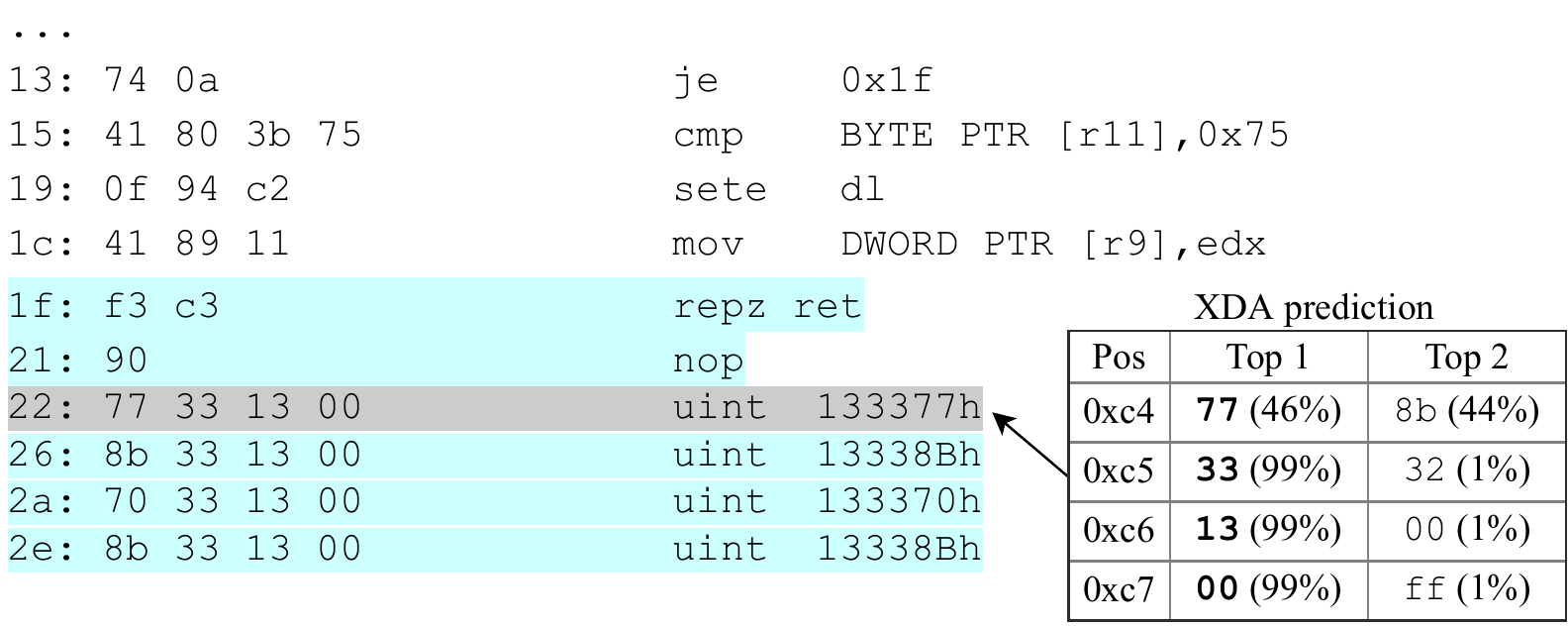}

\caption{We mask a jump table entry which consists of 4 bytes \colorbox{lightgray}{\texttt{77 33 13 00}}. We highlight the function \colorbox{lightblue}{epilogue} and other \colorbox{lightblue}{jump table entries} in the context that help \sys to guess the masked bytes.}
\label{fig:predicting_jump_table_entries}
\end{figure}

\noindent\textbf{Predicting local variable allocation size.}
Consider \texttt{vim} compiled on Windows x64 by Visual Studio with \texttt{O1} in Figure~\ref{fig:predicting_local_variable_allocation}, where we mask out the instruction of decreasing the stack pointer \texttt{rsp} to allocate 40 bytes (\texttt{0x28}) space for local variables. 
At position \texttt{0x24}, we can see the corresponding instruction (\texttt{add rsp, 0x28}) that increases the stack pointer to free the space. While the instruction of increasing and decreasing stack pointer are similar with only the third byte (\texttt{ec} and \texttt{c4}, respectively) different, \sys predicts the masked third byte to be \texttt{ec} with 99\% confidence.
This observation implies that the model is not simply matching similar patterns in its input sequence. Instead, it learns the necessary function prologue/epilogue signature and understands the semantics of underlying instructions, \eg by correctly predicting local variable size \texttt{28} that the program intends to allocate.

\begin{figure}[!t]
\centering

\includegraphics[width=\linewidth]{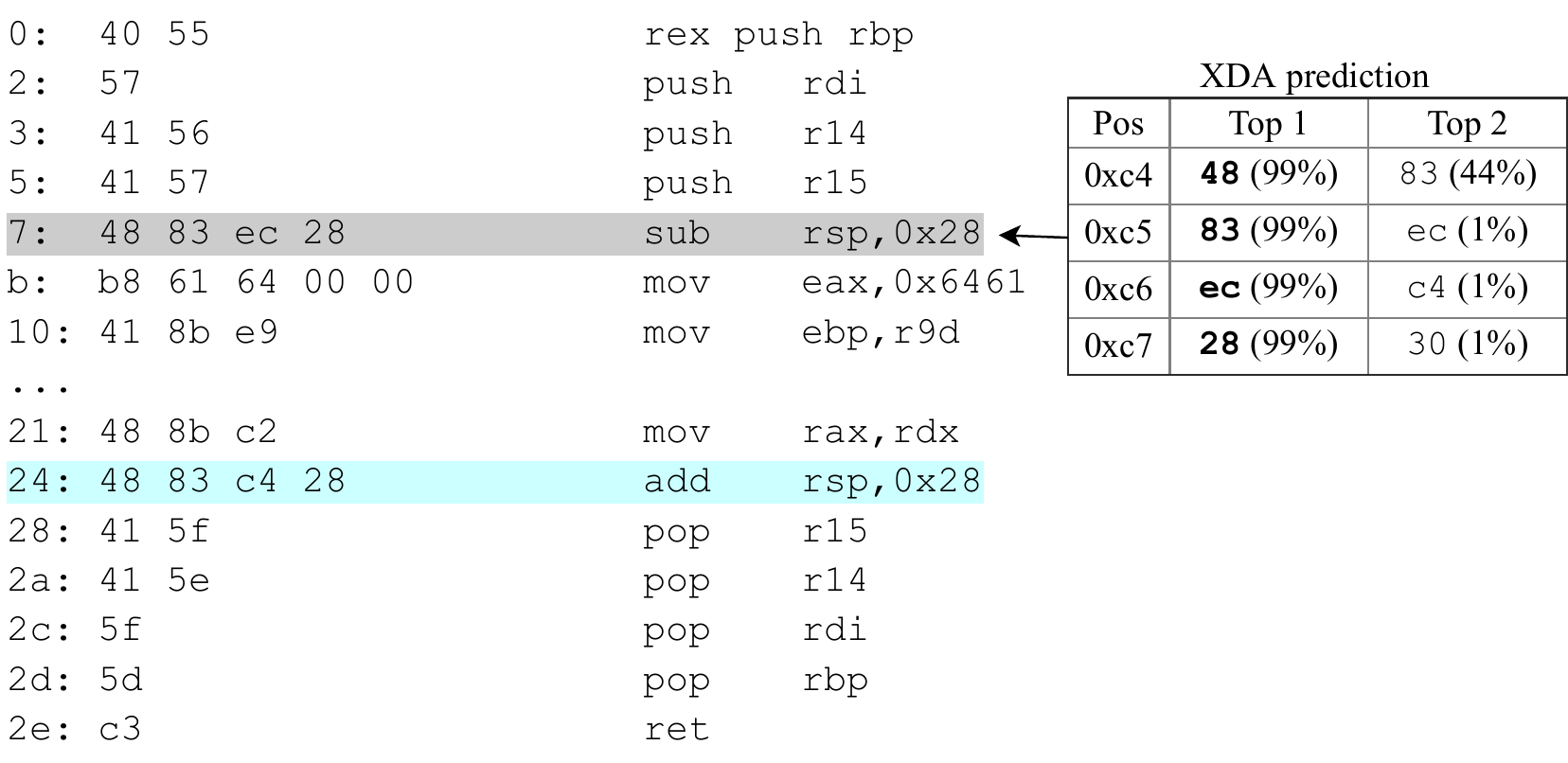}

\caption{We mask the instruction that adjust the stack pointer to allocate \texttt{0x28} (30) bytes on stack for storing the local variable.
It consists of 4 bytes \colorbox{lightgray}{\texttt{48 83 ec 28}}. We mark the instruction that \colorbox{lightblue}{increases the stack pointer} to free the local variable that \sys leverages to make predictions.}
\label{fig:predicting_local_variable_allocation}
\end{figure}

\subsection{Attention Visualization}
\label{subsec:visualization}

We visualize \sys's internal attentions when predicting masked bytes, which explains which input part the model focuses on to make predictions~\cite{wiegreffe2019attention, clark2019what}.
We use the example of allocating local variables described in Section~\ref{sec:overview} to visualize \sys's attention in predicting the (masked) instruction of local variable allocation.
As \sys consists of 12 attention layers with 12 attention heads (detailed in Section~\ref{subsec:pretrain_method}), we take the summation from all layers and attention heads. 

\begin{figure}[!t]
\centering

\includegraphics[width=.85\linewidth]{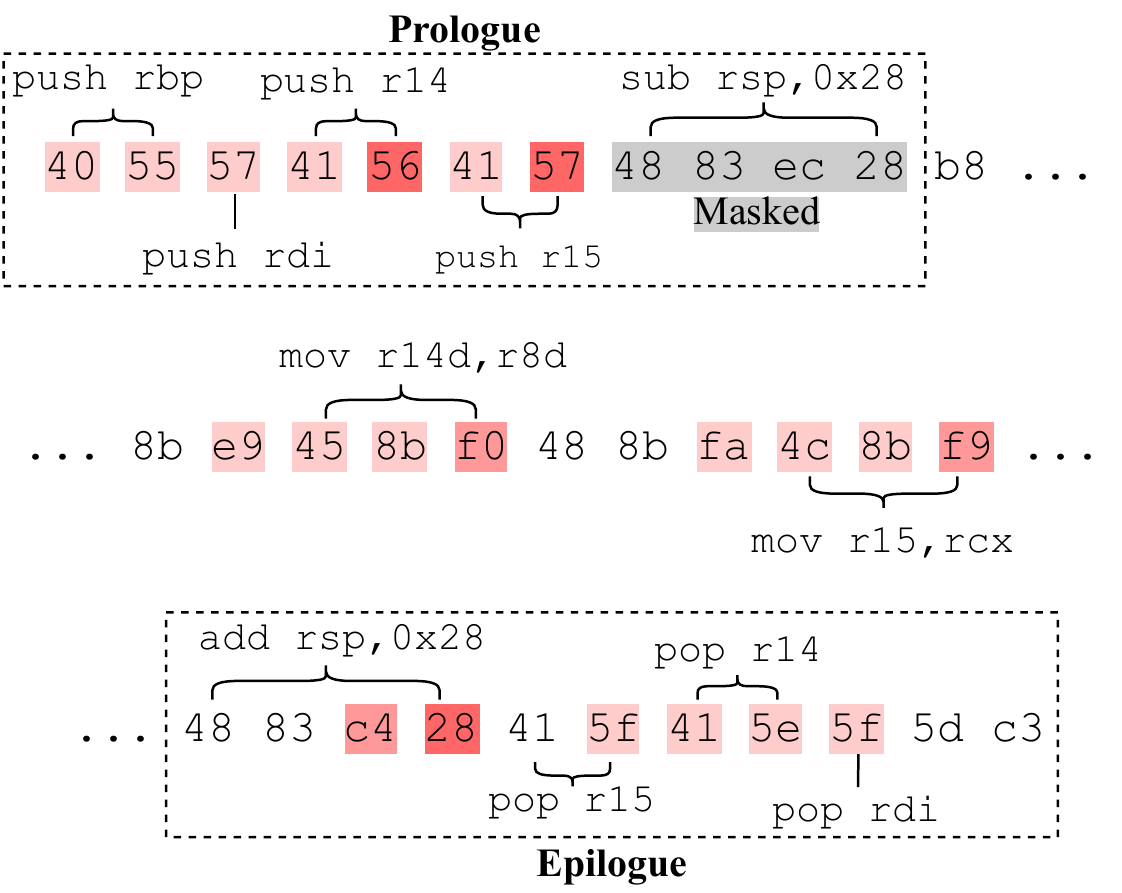}

\caption{The byte sequence example shown in Figure~\ref{fig:ida_birnn_fail} with more bytes in the context. We illustrate the attention distribution of \sys when predicting the masked bytes (the instruction \texttt{sub rsp,0x28}). The darker the color, the larger the attention value. We put the corresponding instructions beside the most attended bytes.}
\label{fig:visualization}
\end{figure}

Figure~\ref{fig:visualization} shows \sys's attention distribution when predicting the masked bytes \texttt{48 83 ec 28}. 
We find that \sys focuses the most on the hints from the prologue and epilogue. 
For example, it notices (with high attention value) the typical argument passing instructions (\texttt{push rdi; push r14; push r14; push r15;}) before the masked bytes, which strongly indicates the following instruction should be decreasing stack pointer (\texttt{sub rsp,-}) for local variable allocation. 
Then it focuses on the last two bytes (\texttt{c4 28}) in the instruction \texttt{add rsp, 0x28}, which increases the stack pointer to free the local variable. 
These two bytes provides the necessary information to decide how many bytes to allocate in the masked instructions. 
Moreover, it also pays attention to some other instructions (\eg \texttt{pop r15; pop r14; pop rdi;}) in the epilogue. This probably further validates that the masked instruction resides within a function body, so that it should leverage the bytes \texttt{c4 28} in the epilogue to predict the masked \texttt{ec 28}. 
Again, note that all instruction, prologue and epilogue information are provided by us for explanation purpose. \sys does not have access to such information but plain bytes. However, the attention visualization strongly implies that \sys implicitly learns all such knowledge. 

\section{Related Work}
\label{sec:related}

Besides works described in Sections~\ref{sec:intro} and~\ref{sec:background}, this section discusses a more comprehensive list of related research papers.


\noindent\textbf{Learned disassembly.}
Statistical learning has long been used for disassembling and binary analysis~\cite{bao2014byteweight, shin2015recognizing, rosenblum2008learning}. 
One major path most ML-based approaches take to boost the accuracy is to incorporate the program semantics.
For example, Wartell~\etal\cite{wartell2011differentiating} leverage the statistical language modeling to capture byte sequence semantics to compress the binaries. Shingled Graph Disassembly~\cite{wartell2014shingled} further employs graph-based learning algorithm with cheaper training cost to distinguish code and inline data. 
Other works~\cite{alrabaee2016bingold, wang2017semantics, miller2019probabilistic} introduce control/data-flow to improve the accuracy and robustness.
\sys employs masked LM as a pretraining step to \emph{automate} learning semantics and transfer to downstream disassembly tasks. 

\noindent\textbf{Applications based on disassembly.}
Disassembly serves as the building block for many security-critical applications.
For example, binary rewriting and hardening~\cite{miller2019probabilistic, williams2020egalito} aims to improve the size, efficiency or security of binaries without  source code access, and rely on disassembly to recover functions as building blocks for rewriting. 
Control-flow integrity~\cite{mohan2015opaque, abadi2009control} and code randomization defenses~\cite{williams2016shuffler} aim to improve the security of binaries by preventing the redirection of control-flow and increasing the difficulty of code injection, respectively. When source code is absent, which is often the case for legacy software in most dire need of such protections, these defenses can be retrofitted to binaries with the help of disassembly and binary rewriting.
Decompilation~\cite{brumley2013native,emmerik2004using} aims to automatically reverse binaries to source code, and relies on disassembly as an initial step to recovering higher-level structure abstracted away during compilation.

\noindent\textbf{Use of ML in other binary analysis tasks.} 
Beyond disassembly, ML has been increasingly applied in other binary analysis tasks.
For example, EKLAVYA~\cite{chua2017neural} learns function type signatures. DeepVSA~\cite{guo2019deepvsa} and RENN~\cite{mu2019renn} learn memory alias dependencies. 
DeepBinDiff~\cite{duan2020deepbindiff} and GEMINI~\cite{xu2017neural} learn function similarity by neural embeddings.
Most of these methods are based on either recurrent networks or graph neural nets that requires nontrivial engineering effort for encoding domain knowledge, which are shown not as effective as \sys's underlying self-attention architecture in capturing long-range dependencies, which is also fully-automated~\cite{vaswani2017attention}.
As shown in Section~\ref{sec:case}, \sys learns much broader knowledge useful beyond the two tasks considered in this paper. 
Therefore, we believe \sys has huge potential in other downstream binary analysis tasks and plan to explore in the future work. 

\section{Conclusion}
\label{sec:conclusion}

We have presented \sys, a novel disassembling technique based on transfer learning. It leverages machine code semantics learned in pretraining masked Language Modeling to solve downstream disassembly tasks accurately, robustly, and efficiently.
\sys is 17.2\% more accurate than the state-of-the-art at recovering function boundaries, and achieves a 99.7\% F1 score at recovering assembly instructions. Furthermore, \sys is robust against various compilers, architectures, platforms, and optimization levels.
Our case studies have shown \sys's potential for a wide range of downstream disassembly and binary analysis tasks beyond recovering functions and instructions.
We open-source \sys at \url{https://github.com/CUMLSec/XDA}.

\section*{Acknowledgment}

We thank our shepherd Kevin Hamlen and the anonymous reviewers for their constructive and valuable feedback. 
This work is sponsored in part by NSF grants CNS-18-42456, CNS18-01426, CNS-16-17670, CNS-16-18771, CCF-16-19123, CNS-15-63843, and CNS-15-64055; ONR grants N00014-17-1-2010, N00014-16-1-2263, and N00014-17-1-2788; an NSF CAREER award; an ARL Young Investigator (YIP) award; a Google Faculty Fellowship; a JP Morgan Faculty Research Award; a DiDi Faculty Research Award; a Google Cloud grant; a Capital One Research Grant; and an Amazon Web Services grant. 
Any opinions, findings, conclusions, or recommendations expressed herein are those of the authors, and do not necessarily reflect those of the US Government, ONR, ARL, NSF, Captital One, Google, JP Morgan, DiDi, or Amazon.


\bibliographystyle{plain}
\bibliography{paper}

\begin{appendix}

\subsection{\sys Hyperparameters}
\label{sec:hyperparm}
We describe the hyperparameters that we used throughout all our experiments, if not mentioned elsewhere explicitly.

\emph{Network architecture.} We use 12 self-attention layers with each having 12 self-attention heads. The embedding dimension is $d_{emb}=768$. We set 3072 as the hidden layer size of MLP $f_{out}$ in the self attention layer. 
Overall, we have roughly 110 million network parameters. 
We adopt GeLU~\cite{hendrycks2016gaussian}, known for addressing the problem of vanishing gradient, as the activation function for \sys's self-attention module. We use the hyperbolic tangent (tanh) as the activation function in finetuning MLP (Equation~\ref{eq:finetune}). 
We set the dropout~\cite{srivastava2014dropout} rate 0.1 for pretraining task while refrain from using dropout in the finetuning task.

\emph{Pretraining and finetuning.} 
We fix the largest input length to be 512 and choose the batch size for both pretraining and finetuning as 8.
For pretraining we use the update frequency as 16, but for finetuning we set as 4. 
The update frequency 16 here means the model will aggregate the gradient for 16 batches before it updates the weight parameter. So pretraining has $16\times 8=128$ effective batch size while finetuning has $4\times 8=32$ effective batch size.
We choose the smaller update frequency for finetuning is because the training data of finetuning is larger due to the extra labels (unlike pretraining where the data is only input bytes). So we restrict loading too much batch size in case running out of GPU memory.
Consequently, to account for the smaller effective batch size, we adopt a relatively smaller learning rate as suggested by the common training tricks~\cite{goyal2017accurate}. We pick $10^{-5}$ for finetuning while pretraining has a larger learning rate $10^{-4}$ as its effective batch size is larger.
Instead of starting with the chosen learning rate at first epoch, we follow the common practice of using small warmup learning rate at first epoch. We use $10^{-7}$ as the initial warmup learning rate, which gets gradually increased until it reaches the actual learning rate after first epoch.
Finally, we use use Adam optimizer, with $\beta_1=0.9$, $\beta_2=0.98$, $\epsilon=10^{-6}$, and weight decay $10^{-2}$. 

In this paper, we adopt the hyperparameters that have shown success in other domains such as sentence entailment in natural language processing~\cite{devlin2018bert,liu2019roberta}. For example, we stack 2-layer fully-connected network for finetuning tasks, because finetuning is often assumed as a simple task as the sophisticated dependency learning is offloaded to the pretraining. While it is possible to search for better hyperparameter choices (\eg more layers or changing connections) from using manual trial~\cite{bergstra2012random} to sophisticated algorithms~\cite{lecun1993automatic}, hyperparameter tuning in general is known as an unsolved task.
We thus leave the comprehensive studies of tuning \sys's hyperparameters in our future work.

\subsection{Common Instructions}

We include the most common instructions or byte sequences, we include the top 5 n-grams of each dataset in Table~\ref{tab:ngram}. We choose $n=1,2,3$ as most instruction types can be denoted within these lengths of bytes~\cite{guide2011intel}. We divide each dataset by the instruction set, as x86 and x86-64 can have highly different mapping rules of disassembling instructions.
We exclude the most common bytes in the results, such as \texttt{cc} (padding bytes used to call to interrupt procedure), and \texttt{00} and \texttt{ff} introduced in two's complement (for addressing). 
There are several interesting patterns emerge in the dataset.
For example, \texttt{8b} (MOV instruction) always ranks among the top.
When \texttt{8b} appears in the 2-gram, the top-ranked is \texttt{8b 45} in x86, which maps to \texttt{MOV *, eax}.
While in x86-64, \texttt{48} appears the most, which is the prefix that specifies the 64 bit operand. Therefore, the top 2-gram in x86-64 is \texttt{48 8b}, which indicates the \texttt{MOV} instruction with 64-bit operand.

\begin{table*}[!t]
\footnotesize
\setlength{\tabcolsep}{4pt}
\centering
\renewcommand{\arraystretch}{1.1}

\caption{Most frequent n-grams (n=1,2,3) for each dataset. We omit the cases that include padding \texttt{cc}, and the common bytes introduced by two's complement \texttt{ff} and \texttt{00}.}
\label{tab:ngram}

\begin{tabular}{c|cc|cc|cc|cc|cccc}
\toprule
\multirow{2}{*}{} & \multicolumn{6}{c|}{x86} & \multicolumn{6}{c}{x86-64} \\
 & 1-gram & count & 2-gram & count & 3-gram & count & 1-gram & count & 2-gram & \multicolumn{1}{c|}{count} & 3-gram & count \\ \midrule 
\multirow{5}{*}{\begin{tabular}[c]{@{}c@{}}SPEC\\ 2017\end{tabular}} & \texttt{8b} & 16,309,260 & \texttt{83 c4} & 4,055,543 & \texttt{83 c4 10} & 2,094,343 & \texttt{48} & 67,406,284 & \texttt{48 8b} & \multicolumn{1}{c|}{23,456,799} & \texttt{20 20 20} & 3,217,114 \\
 & \texttt{83} & 11,315,789 & \texttt{44 24} & 2,747,236 & \texttt{8b 44 24} & 1,079,364 & \texttt{8b} & 35,597,922 &  \texttt{48 89} & \multicolumn{1}{c|}{14,122,963} & \texttt{48 8b 85} & 3,105,839 \\
 & \texttt{24} & 9,407,287 & \texttt{83 ec} & 2,568,435 & \texttt{83 ec 08} & 680,529 & \texttt{89} & 22,438,962 & \texttt{48 8d} & \multicolumn{1}{c|}{7,953,318} & \texttt{48 8b 45} & 2,417,498 \\
 & \texttt{89} & 7,644,124 & \texttt{8b 45} & 2,332,870 & \texttt{89 44 24} & 633,151 & \texttt{24} & 17,730,247 &  \texttt{44 24} & \multicolumn{1}{c|}{4,730,699} & \texttt{48 89 c7} & 1,700,933 \\
 & \texttt{e8} & 7,036,366 & \texttt{c4 10} & 2,095,990 & \texttt{8b 45 08} & 592,854 & \texttt{85} & 14,285,564 &  \texttt{48 83} & \multicolumn{1}{c|}{4,475,293} & \texttt{8b 44 24} & 1,691,308 \\ \midrule
\multirow{5}{*}{\begin{tabular}[c]{@{}c@{}}SPEC\\ 2006\end{tabular}} & \texttt{8b} & 4,411,454 & \texttt{83 c4} & 1,106,643 & \texttt{83 c4 10} & 616,278 & \texttt{48} & 9,048,147 & \texttt{48 8b} & \multicolumn{1}{c|}{3,271,527} & \texttt{48 8b 45} & 383,183 \\
 & \texttt{83} & 3,036,396 & \texttt{83 ec} & 778,944 & \texttt{83 ec 0c} & 376,601 & \texttt{8b} & 5,226,813 & \texttt{48 89} & \multicolumn{1}{c|}{2,224,299} & \texttt{8b 44 24} & 309780 \\
 & \texttt{08} & 2,112,428 & \texttt{8b 45} & 656,627 & \texttt{83 ec 08} & 198,326 & \texttt{89} & 3,690,654 & \texttt{44 24} & \multicolumn{1}{c|}{868,788} & \texttt{48 89 c7} & 305,582 \\
 & \texttt{89} & 2,037,985 & \texttt{c4 10} & 616,663 & \texttt{8b 45 08} & 193,443 & \texttt{24} & 3,152,961 & \texttt{48 83} & \multicolumn{1}{c|}{831,444} & \texttt{89 c7 e8} & 280,042 \\
 & \texttt{e8} & 2,014,742 & \texttt{44 24} & 470,326 & \texttt{8b 44 24} & 162,405 & \texttt{0f} & 2,995,076 & \texttt{48 8d} & \multicolumn{1}{c|}{789,078} & \texttt{89 44 24} & 271,837 \\ \midrule
\multirow{5}{*}{BAP} & \texttt{8b} & 4,859,036 & \texttt{44 24} & 1,428,109 & \texttt{c7 44 24} & 554,131 & \texttt{48} & 7,525,125 & \texttt{48 8b} & \multicolumn{1}{c|}{2,881,802} & \texttt{48 8b 45} & 534,246 \\
 & \texttt{24} & 4,753,064 & \texttt{8b 45} & 1,007,213 & \texttt{89 44 24} & 474,436 & \texttt{8b} & 4,495,772 & \texttt{48 89} & \multicolumn{1}{c|}{1,725,358} & \texttt{8b 44 24} & 233,303 \\
 & \texttt{89} & 3,503,718 & \texttt{24 04} & 572,253 & \texttt{44 24 04} & 385,919 & \texttt{89} & 3,055,089 & \texttt{48 83} & \multicolumn{1}{c|}{703,937} & \texttt{89 44 24} & 214,676 \\
 & \texttt{08} & 2,034,947 & \texttt{c7 44} & 555,754 & \texttt{89 04 24} & 326,154 & \texttt{24} & 2,579,139 & \texttt{44 24} & \multicolumn{1}{c|}{653,945} & \texttt{48 85 c0} & 185,531 \\
 & \texttt{04} & 1,919,547 & \texttt{54 24} & 518,124 & \texttt{8b 44 24} & 270,141 & \texttt{0f} & 1,799,445 & \texttt{8b 45} & \multicolumn{1}{c|}{605,920} & \texttt{48 89 c7} & 180,814 \\ \bottomrule
\end{tabular}
\end{table*}

\subsection{Pretraining on all datasets}
\label{subsec:pretrain_all}

As described in Section~\ref{sec:impl}, we always keep the data used for pretraining, finetuning, and testing strictly separated, to compare with other baselines fairly. 
However, as argued in Section~\ref{subsec:pretrain}, we can always collect a large corpus of binaries in the wild for pretraining, and always reuse the same pretrained model for finetuning on different labeled datasets. Therefore, we have also pretrained a single model using all datasets available in Table~\ref{tab:dataset}.
We then re-run all finetuning experiments of Table~\ref{tab:overall_result}, and include the updated results in Table~\ref{tab:overall_result_pretrain_all}.

\begin{table*}[!t]
\footnotesize
\setlength{\tabcolsep}{7.5pt}
\centering
\renewcommand{\arraystretch}{1.1}

\caption{We update the results of \sys in Table~\ref{tab:overall_result} by using the pretrained model on \emph{all} available datasets. }
\label{tab:overall_result_pretrain_all}
\begin{tabular}{c|c|c||ccccc|ccccc}
\toprule
\multirow{2}{*}{Dataset} & \multirow{2}{*}{Platform} & \multirow{2}{*}{ISA} & \multicolumn{5}{c|}{Recovering Function Boundaries F1 (\%)} & \multicolumn{5}{c}{Recovering Instructions F1 (\%)} \\
 &  &  & \sys & Nucleus & bi-RNN & IDA & Ghidra & \sys & bi-RNN & IDA & Ghidra & objdump \\ \midrule 
\multirow{4}{*}{\begin{tabular}[c]{@{}c@{}}SPEC\\ 2017\end{tabular}} & \multirow{2}{*}{Linux} & x86 & \textbf{98.4} & 55.4 & 79.9 & 91.8 & 89.0 & 99.9 & 87.1 & 95.9 & 94.6 & \textbf{100.0}$^{\dagger}$ \\ \cline{3-13} 
 &  & x64 & \textbf{99.6} & 55.0 & 79.2 & 90.2 & 89.5 & 99.9 & 88.9 & 95.8 & 95.9 & \textbf{100.0}$^{\dagger}$ \\ \cline{2-13} 
 & \multirow{2}{*}{Windows} & x86 & \textbf{99.9} & 60.8 & 73.8 & 67.6 & 70.4 & 99.3 & 82.3 & 96.7 & 92.1 & \textbf{99.3} \\ \cline{3-13} 
 &  & x64 & \textbf{99.9} & 65.0 & 78.4 & 78.0 & 71.6 & \textbf{99.6} & 81.9 & 97.1 & 93.1 & 99.3 \\ \hline
\multirow{4}{*}{\begin{tabular}[c]{@{}c@{}}SPEC\\ 2006\end{tabular}} & \multirow{2}{*}{Linux} & x86 & \textbf{99.8} & 57.2 & 86.7 & 95.7 & 92.2 & 99.9 & 89.0 & 96.3 & 95.5 & \textbf{100.0}$^{\dagger}$ \\ \cline{3-13} 
 &  & x64 & \textbf{99.9} & 56.8 & 73.8 & 92.8 & 92.0 & 99.9 & 85.9 & 96.4 & 94.9 & \textbf{100.0}$^{\dagger}$ \\ \cline{2-13} 
 & \multirow{2}{*}{Windows} & x86 & \textbf{99.9} & 68.2 & 78.5 & 77.9 & 76.3 & \textbf{99.9} & 89.9 & 98.1 & 94.5 & 99.1 \\ \cline{3-13} 
 &  & x64 & \textbf{99.9} & 56.8 & 72.7 & 90.1 & 86.2 & \textbf{99.5} & 86.2 & 97.9 & 95.7 & 99.4 \\ \hline
\multirow{4}{*}{BAP} & \multirow{2}{*}{Linux} & x86 & \textbf{99.6} & 61.5 & 74.1 & 59.0 & 57.2 & N/A$^*$ & N/A$^*$ & N/A$^*$ & N/A$^*$ & N/A$^*$ \\ \cline{3-13} 
 &  & x64  & \textbf{99.7} & 53.5 & 79.0 & 58.3 & 56.5 & N/A$^*$ & N/A$^*$ & N/A$^*$ & N/A$^*$ & N/A$^*$ \\ \cline{2-13} 
 & \multirow{2}{*}{Windows} & x86 & \textbf{99.5} & 69.0 & 80.1 & 89.9 & 87.0 & N/A$^*$ & N/A$^*$ & N/A$^*$ & N/A$^*$ & N/A$^*$ \\ \cline{3-13} 
 &  & x64 & \textbf{99.9} & 70.0 & 81.4 & 90.5 & 80.6 & N/A$^*$ & N/A$^*$ & N/A$^*$ & N/A$^*$ & N/A$^*$ \\ \hline
 \multicolumn{3}{c||}{Average} & \textbf{99.6} & 60.8 & 78.1 & 81.8 & 79.0 & \textbf{99.8} & 86.4 & 96.8 & 94.4 & 99.6 \\ \bottomrule
\multicolumn{13}{l}{\scriptsize $^*$The BAP corpus does not contain source code and PDB files, which are necessary to obtain the assembly instruction ground truth.}\\
\multicolumn{13}{l}{\scriptsize $^{\dagger}$GCC does not generate inline data, so a simple linear disassembler can achieve 100\% F1 score~\cite{andriesse2016depth}. objdump always fails to identify inline data, but because} \\ 
\multicolumn{13}{l}{\scriptsize inline data makes up only a tiny fraction (\textless 1\%) of the code section, objdump's overall F1 score is high.} \\
\end{tabular}
\end{table*}

Table~\ref{tab:overall_result_pretrain_all} shows that the average performance of \sys can be further improved by 0.6\% and 0.1\% for recovering function boundaries and assembly instructions, respectively. 
While in such case, pretraining and finetuning data can overlap, note that pretraining does not train with any labels (\eg function boundaries) but just raw bytes. Indeed, this is the common practice in transfer learning~\cite{dagan2005pascal, rajpurkar2016squad, devlin2018bert, liu2019roberta}.

\end{appendix}

\end{document}